\documentstyle[12pt]{article}
\begin{document}
\date{}

\title{On Nonperturbative Calculations in Quantum Electrodynamics}

\author{V E Rochev\\
{\it Institute for High Energy Physics,}\\{\it Protvino,
Moscow region, Russia}}
\date{}
\maketitle

\newcommand{\be}{\begin{equation}}
\newcommand{\ee}{\end{equation}}
\newcommand{\ba}{\begin{eqnarray}}
\newcommand{\ea}{\end{eqnarray}}
\newcommand{\tr}{\,\mbox{tr}\,}

\begin{abstract}
A new approach to nonperturbative calculations in quantum
electrodynamics is proposed. The approach is based on a regular
iteration scheme for
solution of Schwinger-Dyson equations for generating functional
of Green functions. The approach allows one
to take into account the gauge invariance conditions
(Ward identities) and to perform the renormalization program.

The iteration scheme can be realized in two versions.
The first one ("perturbative vacuum") corresponds to
chain summation in the diagram language. In this version
 the exact result of two-dimensional Schwinger model is reproduced
 at the first step of calculations, but in four-dimensional theory
 the non-physical singularity (Landau pole) arises which leads
 to the triviality of the renormalized theory.
 The second version ("nonperturbative vacuum") corresponds to
 ladder summation and permits one to make  nonperturbative
 calculations of physical quantities in spite of the triviality problem.

For chiral-symmetrical leading approximation two  terms of
 the expansion of the first-step vertex function over photon
 momentum are calculated. The formula $f_2=\alpha/(2\pi-\alpha)$
 for anomalous magnetic moment
is obtained ($\alpha$ is the fine structure constant).

For linearized equation of leading approximation  a problem of
dynamical chiral symmetry breaking is considered, the calculations
are performed for renormalized theory in Minkowski space. In the
strong coupling region $\alpha\ge\pi/3$ the results correspond to
earlier investigations performed in Euclidean theory with cutoff:
solutions arise
 with breakdown of chiral symmetry.
 For the renormalized theory a solution
with breakdown of chiral symmetry is also possible
in the weak coupling region
    $\alpha<\pi/3$, but with a subsidiary condition
    on the value of $\alpha$
    which follows from the gauge invariance.
\end{abstract}

\newpage

\section*{Introduction}

The problem of nonperturbative calculations in quantum
electrodynamics (QED) arose practically simultaneously with the
principal solution of the problem of perturbative calculations
which based on
renormalized coupling constant perturbation theory.
It is necessary to recognize, however, that the progress
in the nonperturbative calculations during last decades
is not too large.
Quantitative description of nonperturbative effects either
is based on non-relativistic foundations ( an example
is the bound state description based on non-relativistic
Coulomb problem)
or is rather open to injury for a criticism.
Besides, the problem of inner inconsistency of QED
exists (see \cite{Bog}). This problem can be formulated as a deep-rooted
thesis on triviality of QED in the nonperturbative
region
(see, for example \cite{Triv}, \cite{Esp} and references therein).

The triviality means that the only non-contradictory
value of the renormalized coupling is zero. Absence of asymptotic
freedom in QED and  unsuccessful looking for
another type of self-consistent ultraviolet behavior are  strong
arguments in favor of the triviality\footnote{ In QED with dynamical
chiral symmetry breaking the situation can be different
 (see \cite{Fomin},\cite{Kosic}).}.
An extremal expression of this point of view is  a statement
that QED can be treated exceptionally
as the renormalized coupling constant perturbation theory,
and, consequently, any nonperturbative calculations are excluded from
the consideration. At a high account such a situation is not desperate
since at very high energies the QED
becomes a part of a grand unification theory which is based on
non-Abelian gauge theory with the self-consistent asymptotically
free ultraviolet behavior.
However, it sounds rather unnatural that for a consistent
relativistic calculation of positronium energy levels, for example,
one should exploit a grand unification theory, while it is not need
for calculations of annihilation cross-sections.
It is difficult to find any physical reasoning for such a principal
difference of these problems. Therefore a possibility to perform
approximate nonperturbative calculations of physical quantities
in the framework of  QED itself without contradictions
with the triviality of the exact theory ( as it made in
the perturbation theory)
seems to be a necessary component of the  theory.

 At first sight the main problem of  nonperturbative calculations
 in QED is  an absence of a universal
 small parameter besides the fine structure constant.
 Due to this reason any
 partial summation of perturbative series seems to be
 an arbitrary procedure
 which can be apologized only by  physical meaning of results.
  At the same time the absence of a small parameter itself
 is not an obstacle  for using some approximation. A motivated
 approximation can give quite satisfactory results even without
 an explicit  small parameter: examples are applications of variational
 methods or mean field type approximations in different areas of
 physical theory.

A general problem of various nonperturbative approximations in QED
is  consistent taking into account of requirements of the gauge invariance
and renormalizability. It is clear that a necessary condition
for such taking into account is the existence of a regular iteration
scheme which in principle permits one to do an arbitrary large amount
of steps toward an exact solution of the problem.
In present work a such type scheme is proposed. The scheme is based
on an approximation of Schwinger-Dyson equations (SDEs) for the generating
functional of Green functions of QED by an exactly soluble equation.
Its solution generates a linear iteration scheme each step of which
is described by a closed system of integro-differential equations.
The requirements of gauge invariance (in the form of  Ward identities)
are easily taken into account at each step of iterations.
The renormalization of equations
of each step is also not a principal problem.
Note that equations for Green functions
at leading approximation and at the first step of iterations
look as familiar ones. Similar equations were written and
investigated earlier in other contexts.
A new thing is its appearance in the structure of the regular
iteration scheme,  and  it is this circumstance that allows one
to give them a successive quantum-field-theoretical interpretation.

Using a bilocal fermion source gives a possibility
to formulate the iteration
scheme in two versions. First of them on the language of Feynman
diagrams of perturbation theory is analog of the summation
of chain diagrams with fermion loop. This version is named
"calculations over perturbative vacuum" since a unique connected
Green function of the leading (vacuum) approximation is the free
electron propagator. The calculations over perturbative vacuum
for two-dimensional electrodynamics give the exact result of
Schwinger model for a photon propagator as early as at the first step,
but for physical four-dimensional case they lead to the appearance
a non-physical Landau pole in
the photon propagator  and,
as a consequence, to the triviality of the renormalized theory.
Thus a practical meaning of the calculations over perturbative
vacuum for four-dimensional QED becomes equal zero, though
they grasp  the main nonperturbative effect --- the triviality
of the full theory.

The second version of the iteration scheme can be compared on the diagram
language with a ladder summation. This version is named  "calculations
over nonperturbative vacuum" since the electron propagator of
the leading vacuum approximation is a solution of a non-trivial
nonlinear equation.
For this version of the iteration scheme the nonperturbative
calculations become possible without contradiction to the
triviality of full theory. The basic part of the work is devoted
to the investigation of this version. Apart from the formulation
of general states and  principles of renormalization of
Green functions, the article contains the calculation
of the first-step vertex function (in the case of
chiral-symmetric solution of the leading approximation equation).
Also, for a linearized version of the theory
the problem of dynamical chiral symmetry
breaking is investigated.

A structure of the paper is the following: in the first Section
the necessary notations and definitions are given; SDEs for the
generating functional of Green functions are introduced in the
formalism of bilocal fermion source, and  generating relations for
Ward identities are considered. In the second Section a general
construction of the iteration scheme for solution of SDEs is
given; also the leading approximation and the first step of
iteration over perturbative vacuum are considered. In Section 3
the renormalization of the first-step equations is made; it is
demonstrated that for the four-dimensional theory the calculations
over perturbative vacuum lead to the trivial theory already at the
first step.
 In Section 4 a scheme of calculations over nonperturbative
vacuum is formulated. In Section 5 two first terms of expansion
of the vertex function in photon momentum are calculated for
chiral-symmetric vacuum. These calculations allow one to obtain a simple
formula for anomalous magnetic moment:
 $f_2=\alpha/(2\pi-\alpha)$,
where $\alpha$ is the fine structure constant.
Section 6 is devoted to investigation of dynamical
chiral symmetry breaking in QED. In Conclusion a discussion
of the results is given.

\section{Schwinger-Dyson equations and Ward identities}

We shall consider a theory of a spinor field
$\psi(x)$ (electron)  interacting with an Abelian gauge field
$A_\mu(x)$ (photon) in
 $n$-dimensional Minkowski space with the metrics
 $x^2\equiv x_\mu x_\mu =
x_0^2-x_1^2-\cdots-x_{n-1}^2$.
(For notation simplicity we write all vector indices as low ones)
The Lagrangian with a gauge fixing term has the form
\be
{\cal L} = -\frac{1}{4}F_{\mu\nu}F_{\mu\nu}
-\frac{1}{2d_l}(\partial_\mu A_\mu)^2
+ \bar\psi(i\hat\partial - m + e\hat A)\psi.
\label{lagrangian}
\ee

Here $F_{\mu\nu}=\partial_\mu A_\nu - \partial_\nu A_\mu,\;
\hat A \equiv A_\mu\gamma_\mu,\; \bar\psi = \psi^*\gamma_0,\;
m$ is an electron mass, $e$ is a charge (coupling constant),
$d_l$ is a gauge parameter,
$\gamma_\mu$ are Dirac matrices.

A generating functional of Green functions (vacuum expectations
values for
$T$-product of fields) can be represented as a functional integral
\be
G(J,\eta) = \int D(\psi,\bar\psi,A)\exp i\Big\{\int dx\Big({\cal L}
+J_\mu(x)A_\mu(x)\Big)
-\int dx dy \bar\psi^\beta(y)\eta^{\beta\alpha}(y,x)\psi^\alpha(x)\Big\}.
\label{G}
\ee

Here $J_\mu(x)$ is a source of the gauge field, and
$\eta^{\beta\alpha}(y,x)$ is a bilocal source of the spinor field
($\alpha$ and $\beta$ are spinor indices). Normalization constant
omitted.

Functional derivatives of
$G$ with respect to sources are vacuum expectation values
\be
\frac{\delta G}{\delta J_\mu(x)} = i<0\mid A_\mu(x)\mid 0>,\;\;
\frac{\delta G}{\delta\eta^{\beta\alpha}(y,x)} =
i<0\mid T\Big\{\psi^\alpha(x)\bar\psi^\beta(y)\Big\}\mid 0>.
\label{VEV}
\ee
A heuristic derivation of SDEs for the generating functional
$G$ is based on relations (see, for example, \cite{Vass},
\cite{Rob})
\be
0 = \int D(\psi,\bar\psi,A)\frac{\delta}{\delta A_\mu(x)}
\exp i\Big\{\int dx\Big({\cal L}+J_\mu(x)
A_\mu(x)\Big)
-\int dx dy \bar\psi(y)\eta(y,x)\psi(x)\Big\},
\label{A}
\ee
\be
0 = \int D(\psi,\bar\psi,A)\frac{\delta}{\delta \bar\psi(x)}\bar\psi(y)
\exp i\Big\{\int dx\Big({\cal L}+J_\mu(x)
A_\mu(x)\Big)
-\int dx dy \bar\psi(y)\eta(y,x)\psi(x)\Big\}.
\label{psi}
\ee
Taking derivatives of eqs.(\ref{A})-(\ref{psi})
and taking into account
eq.(\ref{VEV}) we get SDEs for the generating functional of Green
functions of QED
\be
(g_{\mu\nu}\partial^2-\partial_\mu\partial_\nu+
\frac{1}{d_l}\partial_\mu\partial_\nu)
\frac{1}{i}\frac{\delta G}{\delta J_\nu(x)} +
ie\, \mbox{tr}\Big\{ \gamma_\mu\frac{\delta G}{\delta\eta(x,x)}\Big\}
+ J_\mu(x)G = 0,
\label{SDEA}
\ee
\be
\delta(x-y)G + (i\hat\partial - m)\frac{\delta G}{\delta\eta(y,x)}
+ \frac{e}{i}\gamma_\mu\frac{\delta^2G}{\delta J_\mu(x)
\delta\eta(y,x)} - \int dx' \eta(x,x')
\frac{\delta G}{\delta\eta(y,x')} = 0.
\label{SDEpsi}
\ee
(Here and everywhere below $\partial_\mu$ denote a differentiation
with respect to
variable $x$, while a differentiation with respect to
another variable will be
denoted by  indication of  this variable as an upper index.
For example, the differentiation in variable
$y$ will be denoted as
$\partial_\mu^y$.)

We consider in this and following Sections the unrenormalized
theory, therefore for all singular divergent quantities some
regularization is implied. A renormalization of SDEs is considered
below (Section 3).

Gauge invariance imposes some limitations on solutions of SDEs
(\ref{SDEA}) and (\ref{SDEpsi})  known as
Ward identities.
Acting by $\partial_\mu$ on eq.(\ref{SDEA}), we get a relation
\be
\frac{1}{d_l}\partial^2\partial_\mu\frac{1}{i}\frac{\delta G}{\delta
J_\mu(x)} + e\,\mbox{tr}\,\Big\{i\hat\partial\frac{\delta G}
{\delta\eta(x,x)}\Big\}
+ \partial_\mu J_\mu(x)G = 0,
\label{gen1}
\ee
which we shall name the first generating relation.
One more relation follows from
eq.(\ref{SDEpsi}) and its conjugated SDE
\be
\delta(x-y)G + \frac{\delta G}{\delta\eta(y,x)}(-i\hat\partial^y - m)
+ \frac{e}{i}\frac{\delta^2G}{\delta J_\mu(y)
\delta\eta(y,x)}\gamma_\mu - \int dx'
\frac{\delta G}{\delta\eta(x',x)}\eta(x',y) = 0.
\label{SDEpsicon}
\ee
Eq.(\ref{SDEpsicon}) is a consequence of the relation
\be
0 = \int D(\psi,\bar\psi,A)\frac{\delta}{\delta \psi(y)}\psi(x)
\exp i\Big\{\int dx\Big({\cal L}+J_\mu(x)
A_\mu(x)\Big)
-\int dx dy \bar\psi(y)\eta(y,x)\psi(x)\Big\},
\label{psicon}
\ee
which is conjugated to (\ref{psi}). Subtract eq.(\ref{SDEpsicon})
from eq.(\ref{SDEpsi}), take a trace over spinor indices and put
$y=x$. As a result of these simple manipulations we get the relation
\be
\mbox{tr}\,\Big\{i\hat\partial\frac{\delta G}{\delta\eta(x,x)}\Big\}=
\int dx'\,\mbox{tr}\,\Big\{\eta(x,x')\frac{\delta G}{\delta\eta(x,x')}-
\frac{\delta G}{\delta\eta(x',x)}\eta(x',x)\Big\},
\label{gen2}
\ee
which we shall name the second generating relation.
Relation (\ref{gen2})  is fulfilled really for  numerous class
of interactions local with respect to fermions (see
\cite{NekRo}). For QED we can combine both relation into one,
substituting $\mbox{tr}\,\Big\{i\hat\partial\frac
{\delta G}{\delta\eta(x,x)}\Big\}$
from (\ref{gen2}) into (\ref{gen1}).
As a result we get
\be
\frac{i}{d_l}\partial^2\partial_\mu\frac{\delta G}{\delta
J_\mu(x)} =\partial_\mu J_\mu(x)G+
e\int dx'\,\mbox{tr}\,\Big\{\eta(x,x')\frac{\delta G}
{\delta\eta(x,x')}-
\frac{\delta G}{\delta\eta(x',x)}\eta(x',x)\Big\},
\label{genW}
\ee
which is the generating relation of Ward identities for QED.

Differentiating eq.(\ref{genW}) with respect to
$J_\lambda$ and switching off
the sources we obtain the known relation
\be
\partial^2\partial_\mu D_{\mu\lambda}(x-y)= d_l
\partial_\lambda\delta(x-y),
\label{longA}
\ee
where
\be
D_{\mu\lambda}(x-y)\equiv i\frac{\delta^2G}{\delta J_\lambda(y)
\delta J_\mu(x)}\bigg\vert_{J=\eta=0}
\label{propA}
\ee
is the photon propagator. From relation (\ref{longA})
it follows that the longitudinal part of the full photon propagator
(in momentum space) is as follows:

\be
D_{\mu\lambda}^{long}(k)=-d_l\frac{k_\mu k_\lambda}{(k^2)^2}.
\label{long}
\ee

Differentiating eq.(\ref{genW}) with respect to
$\eta$, we obtain after switching
off the sources another known relation
\be
\frac{i}{d_l}\partial^2\partial_\mu F_\mu(x; x',y')
= e[\delta(x-y')-\delta(x-x')]S(x'-y').
\label{WF}
\ee
Here
\be
F_\mu(x; x',y')\equiv \frac{\delta^2G}{\delta J_\mu(x)
\delta\eta(y',x')}\bigg\vert_{J=\eta=0}
\label{F}
\ee
is the three-point function, and
\be
S(x-y)\equiv \frac{\delta G}{\delta\eta(y,x)}\bigg\vert_{J=\eta=0}
\label{S}
\ee
is the full electron propagator.

Relation (\ref{WF}) looks much familiar if one go to the
amputated three-pointer (vertex function)
$\Gamma_\mu$ which is defined as
\be
\Gamma_\mu(x\mid x',y')
\equiv \int dx_1dx'_1dy'_1S^{-1}(x'-x'_1)F_\nu(x_1; x'_1,y'_1)
S^{-1}(y'_1-y')D^{-1}_{\nu\mu}(x_1-x).
\label{Gamma}
\ee
Exploiting this definition and eq.(\ref{longA}), we obtain
the Ward identity for vertex function
\be
i\partial_\mu \Gamma_\mu(x\mid x',y')=
e[\delta(x-x')-\delta(x-y')]S^{-1}(x'-y').
\label{WGamma}
\ee

\section{Iteration scheme}

At $e=0$ SDEs (\ref{SDEA}) and (\ref{SDEpsi})
have a solution
$$
G^{free} = \exp\Big\{\frac{1}{2i}J_\mu\star D^c_{\mu\nu}\star J_\nu
+ \mbox{Tr}\log(1+S^c\star\eta)\Big\},
$$
where
$$
D^c_{\mu\nu} = [g_{\mu\nu}\partial^2-\partial_\mu\partial_\nu+
\frac{1}{d_l}\partial_\mu\partial_\nu]^{-1},\;\;
S^c = (m-i\hat\partial)^{-1}
$$
are free field propagators, and the sign
$\star$ denotes a multiplication in operator sense.
(Also in operator sense we shall treat
$\mbox{Tr}$, as distinct from
$\mbox{tr}$, which corresponds to the trace over spinor indices.)
The functional $G^{free}$ is a generating functional of Green functions
for free fields and is a base for an iteration scheme of perturbation
theory in coupling constant $e$.

For solution of SDEs
(\ref{SDEA}) and (\ref{SDEpsi}) we shall use  another iteration scheme
 proposed in works \cite{Ro1}, \cite{Ro2}.
A general idea of this scheme is an approximation of
functional-differential SDEs
(\ref{SDEA}) and (\ref{SDEpsi}) by equations with "constant", i.e.
independent of the sources
$J_\mu$  and $\eta$, coefficient. Thus we approximate
functional-differential SDEs near the point
$J_\mu=0,\;\eta=0$. Since the objects of calculations are Green functions,
i.e., derivatives of $G$ in zero, such approximation seems to be quite
natural. A circumstance of no small importance is a simplicity
of the leading approximation of this scheme. Equations for all subsequent
approximations also can be easy written.
Green functions of every order of this scheme are defined as
solutions of a closed system of equations. Technically this scheme
is not much more complicated in comparison with the coupling constant
perturbation theory, but in contrast to the last one, it maintains
an information which is inaccessible in any finite order of the
perturbation theory, i.e. it is a nonperturbative method.

In correspondence with aforesaid, we choose as  leading approximation
equations a system of functional-differential equations
\be
(g_{\mu\nu}\partial^2-\partial_\mu\partial_\nu+
\frac{1}{d_l}\partial_\mu\partial_\nu)
\frac{1}{i}\frac{\delta G^{(0)}}{\delta J_\nu(x)} +
ie\, \mbox{tr}\Big\{ \gamma_\mu\frac{\delta G^{(0)}}{\delta\eta(x,x)}\Big\}
 = 0,
\label{leadA}
\ee
\be
\delta(x-y)G^{(0)} + (i\hat\partial - m)
\frac{\delta G^{(0)}}{\delta\eta(y,x)}
+ \frac{e}{i}\gamma_\mu\frac{\delta^2G^{(0)}}{\delta J_\mu(x)
\delta\eta(y,x)}  = 0.
\label{leadpsi}
\ee
A solution of leading approximation equations (\ref{leadA})-(\ref{leadpsi})
is the functional
\be
G^{(0)}= \exp\Big\{iV_\mu\star J_\mu+\mbox{Tr}\,S^{(0)}\star\eta\Big\},
\label{G0}
\ee
where $V_\mu(x)$ and $S^{(0)}(x,y)$ satisfy equations
\be
(g_{\mu\nu}\partial^2-\partial_\mu\partial_\nu+
\frac{1}{d_l}\partial_\mu\partial_\nu)V_\nu(x)+
ie\,\mbox{tr}\Big\{\gamma_\mu S^{(0)}(x,x)\Big\} =0
\label{charV}
\ee
\be
\delta(x-y)+(i\hat\partial-m)S^{(0)}(x,y) + e\hat V(x)S^{(0)}(x,y)=0.
\label{charS}
\ee
By analogy with the theory of ordinary differential equation we
shall call these equations as characteristic ones.

In correspondence with the choice of the leading approximation
$i$th term of the iteration expansion of the generating functional
\be
G = G^{(0)} + G^{(1)} + \cdots + G^{(i)}+\cdots
\label{Gi}
\ee
is a solution of iteration scheme equations
\be
(g_{\mu\nu}\partial^2-\partial_\mu\partial_\nu+
\frac{1}{d_l}\partial_\mu\partial_\nu)
\frac{1}{i}\frac{\delta G^{(i)}}{\delta J_\nu(x)} +
ie\, \mbox{tr}\Big\{ \gamma_\mu\frac{\delta G^{(i)}}{\delta\eta(x,x)}\Big\}
= -J_\mu(x)G^{(i-1)},
\label{iA}
\ee
\be
\delta(x-y)G^{(i)} +
(i\hat\partial - m)\frac{\delta G^{(i)}}{\delta\eta(y,x)}
+ \frac{e}{i}\gamma_\mu\frac{\delta^2G^{(i)}}{\delta J_\mu(x)
\delta\eta(y,x)} = \int dx' \eta(x,x')
\frac{\delta G^{(i-1)}}{\delta\eta(y,x')}.
\label{ipsi}
\ee
A solution of eqs.(\ref{iA})-(\ref{ipsi}) is looked for in the form
\be
G^{(i)}=P^{(i)}G^{(0)}. \label{P(i)} \ee On taking into account
characteristic equations (\ref{charV})-(\ref{charS}) a system of
equations for $P^{(i)}$ assumes the form
\be
(g_{\mu\nu}\partial^2-\partial_\mu\partial_\nu+
\frac{1}{d_l}\partial_\mu\partial_\nu)
\frac{1}{i}\frac{\delta P^{(i)}}{\delta J_\nu(x)} +
ie\, \mbox{tr}\Big\{ \gamma_\mu\frac{\delta P^{(i)}}{\delta\eta(x,x)}\Big\}
= -J_\mu(x)P^{(i-1)},
\label{PiA}
\ee
\ba
(i\hat\partial - m)\frac{\delta P^{(i)}}{\delta\eta(y,x)}
+ \frac{e}{i}\gamma_\mu\frac{\delta^2P^{(i)}}{\delta J_\mu(x)
\delta\eta(y,x)}+e\hat V(x)\frac{\delta P^{(i)}}{\delta\eta(y,x)}
+ \frac{e}{i}\gamma_\mu S^{(0)}(x,y)\frac{\delta P^{(i)}}{\delta J_\mu(x)}
\nonumber\\
= \int dx' \eta(x,x')
\Big\{\frac{\delta P^{(i-1)}}{\delta\eta(y,x')}
+S^{(0)}(x',y)P^{(i-1)}\Big\}.
\label{Pipsi}
\ea
Since $P^{(0)}\equiv 1$, it is evident that for any $i$
the functional $P^{(i)}$ is a polynomial in functional variables
$J$ and $\eta$. This circumstance is very important  since it means
the system of equations for coefficient functions of this functional
to be closed in any order of the iteration scheme.

This iteration scheme has no explicit small parameter. In some
sense, the sources $J$ and $\eta$  play the role of such a
parameter. Expansion (\ref{Gi}) of the generating functional
should be treated as an approximation of $G(J,\eta)$ near the
point $J_\mu=0,\;\eta=0$.
 In
essence, instead of the question about the small parameter, one
should  put a question about the convergence of the iteration
series. Studying the convergence properties of the expansion for
the quantum-field-theoretical case is obviously a very complicated
problem, and  we have no any rigorous statement concerning this
question. Nevertheless, some qualitative reasons  can be given in
favor of the following supposition: the convergence of such a type
 expansion in any case is not worse than the convergence of the
coupling constant perturbation series. The reason is following:
 being the simplest from the practical point of view the
  coupling constant
perturbation series is the worst in the mathematical sense. The
matter is that a small parameter (the coupling constant) is a
multiplier at the highest derivative terms of the
functional-differential SDEs for the generating functional. This
means that the equation is a singularly perturbed one and,
consequently, the coupling constant perturbation series is an
asymptotic series at best. From the other hand, the proposed
expansion
 is regular in the mathematical sense and, consequently,
possesses the better convergence properties. (See also \cite{Ro2}
for more details.) In addition, note that for a model
"zero-dimensional" theory at $n=0$, for which the functional
integral becomes the ordinary ones, and SDEs --- ordinary
differential equations, such an expansion possesses  very good
convergence properties \cite{Ro2}.

Generating relation (\ref{genW}) of Ward identities in the framework
of given iteration scheme takes  the form
\be
\frac{i}{d_l}\partial^2\partial_\nu\frac{\delta G^{(i)}}{\delta
J_\nu(x)} =\partial_\nu J_\nu(x)G^{(i-1)}+
\int dx'\,\mbox{tr}\,\Big\{\eta(x,x')\frac{\delta G^{(i-1)}}
{\delta\eta(x,x')}-
\frac{\delta G^{(i-1)}}{\delta\eta(x',x)}\eta(x',x)\Big\},
\label{genWi}
\ee
    Its corollaries like eqs.
(\ref{propA}), (\ref{WF}) and (\ref{WGamma}) should be modified
correspondingly. These Ward identities are very useful tools to
control the gauge invariance. As we can see below, the gauge
invariance requirements, which are  expressed in the form of Ward
identities, impose some strict limitations in given iteration
scheme (see, for example, discussions after eq. (\ref{S(1)r}) and
in Section 5). First step calculations, which are considered in
present article, demonstrate their compatibility with the gauge
invariance (though some question arises for the solutions with
chiral symmetry breaking, see Sections 6 and Conclusion). Surely,
at the moment we can say nothing on this problem for higher steps
of iterations.

Consider in more detail the leading approximation described by eqs.
(\ref{G0}), (\ref{charV}) and (\ref{charS}).
It is evident that for a conservation of Poincar\'e-invariance of the
theory one should assume
$V_\mu\equiv 0$.
Then from eq.(\ref{charV}) it follows that
\be
\mbox{tr}\,\Big\{\gamma_\mu S^{(0)}(x,x)\Big\}=0.
\label{condS}
\ee
By definition (see (\ref{S})) $S^{(0)}$ is the electron propagator
at the leading approximation, therefore it follows from above mentioned
Poincar\'e-invariance that
$S^{(0)}(x,y)=S^{(0)}(x-y)$,
and
$$
S^{(0)}(x,x)=S^{(0)}(0)= \int \frac{dp}{(2\pi)^n}S^{(0)}(p).
$$
As was stated above, a regularization is always supposed for all such
like expressions.

As these conditions being fulfilled the solution of eq.(\ref{charS})
is the free propagator
$$S^{(0)}=S^c=(m-i\hat\partial)^{-1}.$$
For this solution  condition (\ref{condS})
is equivalent to the requirement imposed on a regularization
$$
\int dp\frac{p_\mu}{m^2-p^2}=0.
$$
It is very difficult to imagine an invariant regularization
for which this condition would not be fulfilled.

So, the generating functional of leading approximation is
\be
G^{(0)}= \exp\Big\{\mbox{Tr}\,S^á\star\eta\Big\}.
\label{G0p}
\ee

As  follows from eq.(\ref{G0p}),
the unique connected Green function of the leading approximation
is the electron propagator. Other connected Green functions  appear
at following iteration steps. It is also follows from eq.(\ref{G0p})
that the generating functional at leading approximation  does not possess
the complete Fermi-symmetry.
Really, as  follows from the definition of
generating functional, the Fermi-symmetry implies on {\it full}
generating functional the condition
\be
\frac{\delta^2G}{\delta\eta^{\beta\alpha}(y,x)
\delta\eta^{\beta'\alpha'}(y',x')}=
-\frac{\delta^2G}{\delta\eta^{\beta'\alpha}(y',x)
\delta\eta^{\beta\alpha'}(y,x')}.
\label{fermi}
\ee
Evidently  condition
(\ref{fermi}) does not fulfilled for $G^{(0)}$ defined by eq.(\ref{G0p}).
The violation of this condition
leads particularly to the violation of connected structure
of the leading approximation two-particle (four-point) electron
function which is
\be
S_2^{(0)}(x,y;x',y')\equiv \frac{\delta^2G^{(0)}}
{\delta\eta^{\beta\alpha}(y,x)
\delta\eta^{\beta'\alpha'}(y',x')}\bigg\vert_{J=\eta=0}=
S^c(x-y)S^c(x'-y')
\label{S20}
\ee
-- a term $-S^c(x-y')S^c(x'-y)$ is missed.

Such a situation is rather typical for nonperturbative calculational
schemes with  bilocal source (for example, for
$1/N$-expansion in the bilocal source formalism),
but discrepancy of  such type are not an obstacle for using
 these iteration schemes. Really, condition
(\ref{fermi}) should be satisfied by the {\it full} generating
functional $G$ which is an {\it exact} solution of SDEs
(\ref{SDEA}) and (\ref{SDEpsi}). It is clear that an approximate
solution may do not possess all properties of an exact one. In
given case we have  just the same situation. Properties of
connectivity and Fermi-symmetry of higher Green functions, which
are not fulfilled at first steps of the iteration scheme, restore
at subsequent steps. For example, the structure of disconnected
part of the two-electron function is restored as early as at the
first step of the iteration scheme (see below). At subsequent
steps the correct connected structure of many-electron functions
and other corollaries of Fermi-symmetry are reconstructed. Such
stepwise reconstruction of exact solution properties is very
natural for the given iteration scheme as  it is based on an
approximation of the generating functional $G(J,\eta)$ in vicinity
of zero. The Green functions are coefficients of the generating
functional expansion in the vicinity of zero, therefore only the
lowest functions are well-described at first steps of the
approximation -- at the leading approximation the electron
propagator only. Higher many-particle functions come into the play
later, at following steps, and relation  (\ref{fermi}) is
fulfilled more and more exactly when we go toward exact solution.

A solution of the first step equation is looked for in the form
$$
G^{(1)}= P^{(1)}G^{(0)},
$$
where polynomial $P^{(1)}$ is a solution of eqs.(\ref{PiA}) and
(\ref{Pipsi})
at $i=1$ and $V_\mu\equiv 0$. This solution is
\be
P^{(1)}= \frac{1}{2}\eta\star S^{(1)}_2\star\eta +
S^{(1)}\star\eta + \frac{1}{2i}J_\mu\star D^{(1)}_{\mu\nu}\star J_\nu
+ J_\mu\star F_\mu^{(1)}\star\eta + i{\cal A}^{(1)}_\mu\star J_\mu.
\label{P1}
\ee
In correspondence with above definitions,
$S^{(1)}_2(x,y;x',y')$ is a two-electron function,\\
$S^{(1)}(x-y)$ is a correction to the electron propagator,
$D^{(1)}_{\mu\nu}(x-y)$ is a photon propagator,
 $F_\mu^{(1)}(x; x',y')$ is a three-point function,
 ${\cal A}^{(1)}_\mu$ is a photon-field vacuum expectation value.
Upper index denotes the first-step iteration scheme quantities.

Equations for first-step functions follow from eqs.(\ref{PiA}),(\ref{Pipsi})
at $i=1$, $V_\mu\equiv 0$ and they have the form
\be
(g_{\mu\nu}\partial^2-\partial_\mu\partial_\nu+
\frac{1}{d_l}\partial_\mu\partial_\nu)D^{(1)}_{\nu\lambda}(x-y)-
ie\tr\Big\{\gamma_\mu F^{(1)}_\lambda(y\mid x,x)\Big\}=
g_{\mu\lambda}\delta(x-y),
\label{DF}
\ee
\be
(g_{\mu\nu}\partial^2-\partial_\mu\partial_\nu+
\frac{1}{d_l}\partial_\mu\partial_\nu)F^{(1)}_\nu(x\mid x',y')=
e\tr\Big\{\gamma_\mu S^{(1)}_2(x,x;x',y')\Big\},
\label{FS2}
\ee
\be
(g_{\mu\nu}\partial^2-\partial_\mu\partial_\nu+
\frac{1}{d_l}\partial_\mu\partial_\nu){\cal A}^{(1)}_\nu(x)+
ie\tr \Big\{\gamma_\mu S^{(1)}(x,x)\Big\}=0,
\label{AS}
\ee
\be
(i\hat\partial-m)S^{(1)}_2(x,y;x',y')-ie\gamma_\mu S^c(x-y)
F^{(1)}_\mu(x\mid x',y')=\delta(x-y')S^c(x'-y),
\label{S2F}
\ee
\be
(i\hat\partial-m)F^{(1)}_\lambda(z\mid x,y)=e\gamma_\mu S^c(x-y)
D^{(1)}_{\mu\lambda}(x-z),
\label{FD}
\ee
\be
(i\hat\partial-m)S^{(1)}(x-y)-ie\gamma_\mu F^{(1)}_\mu(x\mid x,y)
+e\gamma_\mu S^c(x-y){\cal A}^{(1)}_\mu(x)=0.
\label{SF}
\ee
The system of eqs.(\ref{DF})-(\ref{SF}) at first sight  seems to be
overfilled: six equations for five functions. Really one can prove
that eq.(\ref{FS2}) is a corollary of other equations.
Further, if one imposes on
$S^{(1)}$ the same condition as for $S^{(0)}$  (see (\ref{condS})),
then  the existence of  trivial solution    ${\cal A}^{(1)}_\mu\equiv 0$
for ${\cal A}^{(1)}_\mu$ follows from eq.(\ref{AS}).
We restrict ourselves to this solution for
${\cal A}^{(1)}_\mu$ ignoring all others as violating Poincar\'e-invariance
of the theory.

From eq.(\ref{FD}) we obtain
\be
F^{(1)}_\lambda(z\mid x,y)=-e\int dx'S^c(x-x')
\gamma_\mu S^c(x'-y)D^{(1)}_{\mu\lambda}(x'-z).
\label{F1}
\ee
Substituting (\ref{F1}) into (\ref{DF}), we obtain after simple
transformations
\be
(D^{(1)}_{\mu\nu})^{-1}=(D^{c}_{\mu\nu})^{-1}+ \Pi_{\mu\nu}
\label{D1}
\ee
where
\be
\Pi_{\mu\nu}(x)=ie^2\tr\Big\{\gamma_\mu S^c(x)\gamma_\nu S^c(-x)\Big\}
\label{loop}
\ee
is a free-electron loop. It follows from Ward identities
that $\Pi_{\mu\nu}$ is transversal in the momentum space:
\be
\Pi_{\mu\nu}(k)=\Pi(k^2)\,\pi_{\mu\nu},
\label{pi}
\ee
where
$$
\pi_{\mu\nu}\equiv
g_{\mu\nu}-\frac{k_\mu k\nu}{k^2}
$$
is transverse projector. On taking into account eq.
(\ref{pi}) we finally obtain (in the momentum space)
\be
D^{(1)}_{\mu\nu}(k)=\frac{1}{-k^2+\Pi(k^2)}\,\pi_{\mu\nu}-
d_l\frac{k_\mu k_\nu}{(k^2)^2}.
\label{D1k}
\ee

Then we obtain from eq.
(\ref{S2F}) a two-electron function
\ba
S^{(1)}_2(x,y;x',y') = -S^c(x-y')S^c(x'-y)+\nonumber\\
+ie^2\int dx_1dx_2S^c(x-x_1)\gamma_\mu S^c(x_1-y)D^{(1)}_{\mu\nu}(x_1-x_2)
S^c(x'-x_2)\gamma_\nu S^c(x_2-y'),
\label{S21}
\ea
and from eq.
(\ref{SF})-- a correction to the electron propagator
\be
S^{(1)}(x-y)= ie^2\int dx_1dx_2S^c(x-x_1)\gamma_\mu
S^c(x_1-x_2)D^{(1)}_{\mu\nu}(x_1-x_2)
\gamma_\nu S^c(x_2-y).
\label{S1}
\ee

Thus we have obtained the expressions for all first-step functions.
Note, that a disconnected part of function
$S^{(1)}_2$ (see (\ref{S21}))
is the "missing" disconnected part of the leading-approximation
two-electron function (\ref{S20}). Hence, as  was mentioned earlier,
the correct structure of the disconnected part of the two-electron
function is reconstructed at the first step of the calculations.
At subsequent steps a correct crossing-symmetrical structure
of the connected part is also reconstructed.

Before  renormalization the formulae obtained are formal
expressions due to ultraviolet divergences. The question arises:
is there an explicit regularization scheme, which is consistent
with the considered iterative expansion?  Though in general case
we have no positive answer for this question for any finite step
of iteration, at the same time we have no reasons to suppose the
usual well-known regularization schemes do not work in our case.
In particular, the ultraviolet divergence of the free-electron
loop (\ref{loop}) can be treated with dimensional regularization.
Note, that at $n=2$ (two-dimensional electrodynamics) this loop
converges in dimensional regularization. At $m=0$ (Schwinger
model)
\be
\Pi(k^2)=\frac{e^2}{\pi},
\label{pi2}
\ee
and the photon propagator (in transverse gauge) is
\be
D^{(1)}_{\mu\nu}(k)=\frac{1}{\frac{e^2}{\pi}-k^2}\,\pi_{\mu\nu},
\label{D1k2}
\ee
which coincides with the exact result \cite{Schw}.

\section{Renormalization}

So far we have considered unrenormalized theory.  To  give a physical
sense to quantities to be calculated in a theory with ultraviolet
divergences it is necessary to carry out the renormalization procedure.

The Lagrangian of QED for the renormalized fields has the form
\be
{\cal L}_r = -\frac{Z_3}{4}F_{\mu\nu}F_{\mu\nu}
-\frac{1}{2d_l}(\partial_\mu A_\mu)^2
+ Z_1\bar\psi(i\hat\partial - m_r + e_r\hat A)\psi - \delta m\bar\psi\psi.
\label{lagrangianr}
\ee
Here $\psi$ and $A_\mu$ are the renormalized fields,
$Z_1$ and $Z_3$ are  renormalization constants of
spinor and gauge fields correspondingly,
 $m_r$ and  $e_r$ are renormalized mass and charge,
$\delta m$ is the counterterm of electron-mass renormalization.
We have taken into account the fact, that due to the gauge invariance
and Ward identities the renormalization constants of the spinor field
and of the  interaction are equal:
$Z_1=Z_2$, and the longitudinal part of the gauge field is not
renormalized.

SDEs for a generating functional of renormalized Green functions are
\be
[Z_3(g_{\mu\nu}\partial^2-\partial_\mu\partial_\nu)+
\frac{1}{d_l}\partial_\mu\partial_\nu]
\frac{1}{i}\frac{\delta G}{\delta J_\nu(x)} +
ie_rZ_1\,\mbox{tr}\,\Big\{ \gamma_\mu\frac{\delta G}{\delta\eta(x,x)}\Big\}
+ J_\mu(x)G = 0,
\label{SDEAr}
\ee
\ba
\delta(x-y)G + Z_1(i\hat\partial - m_r)\frac{\delta G}{\delta\eta(y,x)}
-\delta m\frac{\delta G}{\delta\eta(y,x)}
+ \frac{e_rZ_1}{i}\gamma_\mu\frac{\delta^2G}{\delta J_\mu(x)
\delta\eta(y,x)} =\nonumber\\
=\int dx' \eta(x,x')
\frac{\delta G}{\delta\eta(y,x')}.
\label{SDEpsir}
\ea

Generating relation (\ref{genW}) of Ward identities for the
renormalized generating functional preserves its form under
substitution
$e\rightarrow e_r$. Correspondingly, its  corollaries
(\ref{longA}), (\ref{WF}) and (\ref{WGamma}) are preserved with
the following distinction: all quantities are renormalized ones.

Being applied to our iteration scheme, the renormalization procedure
means: the  generating functional   $G$ as well as
all renormalization constants
and counterterms has corresponding iteration expansions:
$$
Z_j=Z_j^{(0)}+Z_j^{(1)}+\cdots,\;\;\;
\delta m= \delta m^{(0)}+\delta m^{(1)}+\cdots
$$
Thus renormalized equations of leading approximation are
\be
[Z_3^{(0)}(g_{\mu\nu}\partial^2-\partial_\mu\partial_\nu)+
\frac{1}{d_l}\partial_\mu\partial_\nu]
\frac{1}{i}\frac{\delta G^{(0)}}{\delta J_\nu(x)} +
ie_rZ_1^{(0)}\,\mbox{tr}\,\Big\{ \gamma_\mu\frac{\delta G^{(0)}}
{\delta\eta(x,x)}\Big\}=0,
\label{leadAr}
\ee
\be
\delta(x-y)G^{(0)} + Z_1^{(0)}
(i\hat\partial - m_r)\frac{\delta G^{(0)}}{\delta\eta(y,x)}
-\delta m^{(0)}\frac{\delta G^{(0)}}{\delta\eta(y,x)}
+ \frac{e_rZ_1^{(0)}}{i}\gamma_\mu\frac{\delta^2G^{(0)}}{\delta J_\mu(x)
\delta\eta(y,x)} =0.
\label{leadpsir}
\ee
A solution of these equations is the same leading approximation functional
(\ref{G0p}) but now $S^c$ is the renormalized free propagator
$$
S^c=(m_r-i\hat\partial)^{-1}.
$$
At that
$$
Z_1^{(0)}=1,\;\;\delta m^{(0)}=0,
$$
and the same condition (\ref{condS}) is imposed for a regularization.
Note that the value of constant
$Z_3^{(0)}$ remains  undefined in the framework
of leading approximation. This constant is fixed at the first step of
iteration scheme when the photon propagator comes into play.

The first-step functional is $G^{(1)}=P^{(1)}G^{(0)}$, and
taking into account above definitions and results of the
leading approximations we obtain for
$P^{(1)}$  the equations
\be
[Z_3^{(0)}(g_{\mu\nu}\partial^2-\partial_\mu\partial_\nu)+
\frac{1}{d_l}\partial_\mu\partial_\nu]
\frac{1}{i}\frac{\delta P^{(1)}}{\delta J_\nu(x)} +
ie_r\,\mbox{tr}\,\Big\{ \gamma_\mu\frac{\delta P^{(1)}}
{\delta\eta(x,x)}\Big\}=-J_\mu(x),
\label{P1Ar}
\ee
\ba
(i\hat\partial - m_r)\frac{\delta P^{(1)}}{\delta\eta(y,x)}
+ \frac{e_r}{i}\gamma_\mu\frac{\delta^2P^{(1)}}{\delta J_\mu(x)
\delta\eta(y,x)}+
\frac{e_r}{i}\gamma_\mu S^c(x-y)\frac{\delta P^{(1)}}{\delta J_\mu(x)}=
\nonumber\\
= \int dx_1\eta(x,x_1)S^c(x_1-y) + Z_1^{(1)}\delta(x-y) +
\delta m^{(1)}S^c(x-y).
\label{P1psir}
\ea
A solution of these equations has the same form
(\ref{P1})  with a distinction: the coefficient functions
of polynomial (\ref{P1}), which define first-step Green functions,
are now the renormalized quantities. We do not write out equations
for the coefficient functions which coincides with above unrenormalized
equations (\ref{DF})-(\ref{SF}) up to evident from
(\ref{P1Ar}) and (\ref{P1psir}) modifications.

For three-pointer $F^{(1)}_\mu$ we obtain the same expression (\ref{F1}),
and for the photon propagator in the momentum space we have
\be
D^{(1)}_{\mu\nu}(k)=\frac{1}{-Z_3^{(0)}k^2+\Pi(k^2)}\,\pi_{\mu\nu}-
d_l\frac{k_\mu k_\nu}{(k^2)^2},
\label{D1kr}
\ee
where $\Pi$ is the same free-electron loop (\ref{loop})-(\ref{pi})
with a substitution $e\rightarrow e_r\;m\rightarrow m_r$.

Expression (\ref{S21}) for two-electron function $S_2^{(1)}$
is not changed (with the same substitution again). For the first-step
correction to the electron propagator we have
\ba
S^{(1)}(x-y)= ie^2_r\int dx_1dx_2S^c(x-x_1)\gamma_\mu
S^c(x_1-x_2)D^{(1)}_{\mu\nu}(x_1-x_2)
\gamma_\nu S^c(x_2-y)
\nonumber\\
+ Z_1^{(1)}S^c(x-y)-\delta m^{(1)}
\int dx_1S^c(x-x_1)S^c(x_1-y).
\label{S1r}
\ea

If photon propagator (\ref{D1kr}) possesses a pole
in the point $k^2=\mu^2$ which corresponds to a particle
with mass $\mu$, then for $\Pi(k^2)$ the following normalization conditions
should be satisfied
\be
\Pi(\mu^2)=Z_3^{(0)}\mu^2,\;\;\;\Pi'(\mu^2)=Z_3^{(0)}-1.
\label{normA}
\ee
In two-dimensional case ($n=2$)
\be
\Pi(k^2)=-\frac{e^2_rk^2}{\pi}\int_{0}^{1} dx\frac{x(1-x)}
{m^2_r-x(1-x)k^2-i0}.
\label{pi2gen}
\ee
 In particular at $m_r=0:\;\Pi(k^2)=e^2_r/\pi$, and
normalization conditions (\ref{normA}) give
$Z_3^{(0)}=1,\,\mu^2=e_r^2/\pi$,
 i.e. the result for Schwinger model is unchanged by the renormalization.

For four-dimensional theory $\Pi(k^2)$ is an ultraviolet-divergent
quantity. At the dimensional regularization ($n=4-2\epsilon$)
\be
\Pi(k^2)=-\frac{e^2_rk^2}{2\pi^2}(2\pi)^\epsilon\Gamma(\epsilon)
\int_{0}^{1} dx\frac{x(1-x)}
{(m^2_r-x(1-x)k^2-i0)^\epsilon}.
\label{pi4}
\ee
Here $\Gamma(x)$ is gamma-function.
In this case the normalization conditions give us the following
 $$
 Z_3^{(0)}=1-\frac{\alpha_r}{3\pi}\Big\{\frac{1}{\epsilon}+
 \psi(1)+\log 2\pi - \log\frac{m_r^2}{M^2}\Big\},\;\;\;\mu^2=0.
 $$
 Here $\alpha_r=e^2_r/4\pi$,
 $\psi(1)$ is Euler constant, $M^2$ is a mass parameter
 of the dimensional regularization.

 In the Euclidean region $k^2<0$ the renormalized photon propagator
in four-dimensional theory possesses a non-physical pole at
 $k^2\simeq -m^2_r\exp\{\frac{3\pi}{\alpha_r}\}$
 -- it is  Landau pole \cite{Landau}.
 This pole arises also at renormalization-group summation \cite{Bog},
 and in the framework of  $1/N$-expansion \cite{Kir}.
 A presence of this pole is a serious difficulty and, in particular,
 prevents one from making  sensible calculations (or  needs some refined
 construction which seems to be  superfluous for QED \cite{Esp}).
 In the coupling constant perturbation theory one can exploit
 a smallness of $\alpha_r$ to avoid the  troubles, but in
 used nonperturbative scheme the unique non-contradictory possibility
 is the choice  $\alpha_r=0$, and the theory becomes trivial.

 Therefore, as a result of investigation of the first-step equation
 of our calculational scheme the following conclusion  is made:
 for the four-dimensional QED the renormalization of theory leads to
 triviality as early as at the first step of calculations.
 This result, from one side, demonstrates that proposed scheme
 correctly catches a nonperturbative content of QED, but,
from the other side, such a version of the scheme is practically useless
in contrast to the usual perturbation theory.
However, the other version of the same calculational scheme exists
which permits one to make sensible nonperturbative calculations in QED.
To state this version is a  matter of the remainder of the paper.

 \section{Nonperturbative vacuum}

 The iteration scheme considered above can be modified in such a manner
 that it becomes insensitive to triviality at least for first
 nontrivial steps of the calculations, i.e. applicable
 to nonperturbative calculations in QED.
 It means a possibility to calculate physical quantities without
 contradiction with the triviality of  theory, in a similar way it is made
 in the perturbation theory in renormalized coupling constant. Of
 course, one  cannot say that the situation is fully analogous
 with coupling constant perturbation theory. In the framework of
 the perturbation theory one can be sure that calculations in any
 finite order do not lead to pathologies. It would be untimely to allege
 the same statement for the given iteration scheme. One cannot exclude at
 the moment the possibility of a pathology  (which originates from
 the triviality of four-dimensional QED) in a higher step of the iterations.
 In the case of such pathology, the use of a grand unification for
 its elimination seems to be inevitable. A full answer on these
 principal question can be given only with detail investigations of
 higher steps of iterations. Nevertheless, as we shall see, an analysis
 of first steps of given iteration scheme demonstrates that it
 includes in a natural way such well-known nonperturbative
 treatments of QED as ladder and rainbow approximations.

To pass  this modification of the iteration scheme (which we shall
name calculations over nonperturbative vacuum) let us resolve SDE
  (\ref{SDEA}) with regard to the first derivative of the generating
  functional with respect  $J_\mu$:
  \be
  \frac{1}{i}\frac{\delta G}{\delta J_\mu(x)}=
  -\int dx_1 D^c_{\mu\nu}(x-x_1)\Big\{J_\nu(x_1)G+
  ie\tr \gamma_\nu\frac{\delta G}{\delta\eta(x_1,x_1)}\Big\},
  \label{dG/dJ}
  \ee
and substitute it into the second SDE (\ref{SDEpsi}).
As a result we obtain the "integrated over  $A_\mu$"
(in the functional-integral terminology) equation
\ba
\delta(x-y)G + (i\hat\partial - m)\frac{\delta G}{\delta\eta(y,x)}
+ \frac{e^2}{i}
\int dx_1 D^c_{\mu\nu}(x-x_1)\gamma_\mu
\frac{\delta}{\delta\eta(y,x)}\tr \gamma_\nu\frac{\delta G}
{\delta\eta(x_1,x_1)}=
\nonumber\\
= \int dx_1 \Big\{\eta(x,x_1)
\frac{\delta G}{\delta\eta(y,x_1)} +
e D^c_{\mu\nu}(x-x_1)J_\nu(x_1)\gamma_\mu
\frac{\delta G}{\delta\eta(y,x)}\Big\}.
\label{SDE}
\ea
Exploiting Fermi-symmetry condition (\ref{fermi})
let us transform eq.(\ref{SDE}) in following manner:
 \ba
\delta(x-y)G + (i\hat\partial - m)\frac{\delta G}{\delta\eta(y,x)}
+ ie^2
\int dx_1 D^c_{\mu\nu}(x-x_1)\gamma_\mu
\frac{\delta}{\delta\eta(x_1,x)}\gamma_\nu\frac{\delta G}
{\delta\eta(y,x_1)}=
\nonumber\\
= \int dx_1 \Big\{\eta(x,x_1)
\frac{\delta G}{\delta\eta(y,x_1)} +
e D^c_{\mu\nu}(x-x_1)J_\nu(x_1)\gamma_\mu
\frac{\delta G}{\delta\eta(y,x)}\Big\}.
\label{SDEF}
\ea
From the point of view of {\it exact} solutions equations (\ref{SDE})
and (\ref{SDEF}) are fully equivalent since the transition from
eq.(\ref{SDE}) to eq.(\ref{SDEF}) is, in essence, an
identical transformation. However, it is not the case for the
used iteration scheme since, as was said earlier, condition
(\ref{fermi})
is fulfilled  only approximately at any finite step of the iteration scheme.
Therefore, eqs.
(\ref{SDE}) and (\ref{SDEF}) lead to different  expansions.
Eq.(\ref{SDE}) leads, in essence, to the same calculational scheme as above,
but  eq.(\ref{SDEF}) gives us a fruitful  scheme of calculation of
physical quantities.

Before  the writing  equations of the scheme note that the photon
propagator is not contained directly in coefficient functions
of polynomials $P^{(i)}$ of the iteration scheme for eqs.
(\ref{SDE}) and (\ref{SDEF}) since these equations are, as pointed above,
the result of "integration" over the photon variable $A_\mu$.
For  calculation of the photon propagator and concerning quantities
one should exploit the Dyson formula
\be
D_{\mu\nu}(x-y)\equiv
i\frac{\delta^2G}{\delta J_\nu(y)\delta J_\mu(x)}\bigg\vert_{J=\eta=0}
 = D^c_{\mu\nu}(x-y)+ie\int dx_1 D^c_{\mu\rho}(x-x_1)\tr
 \gamma_\rho F_\nu(y;x_1,x_1),
\ee
which is a consequence of a single differentiation of eq.(\ref{dG/dJ}).

In correspondence  with the general principle of iteration construction
we choose as the leading approximation  eqs. (\ref{SDE}) and (\ref{SDEF})
with $J_\mu=0,\,\eta=0$ in their coefficients. These equations are
\be
\delta(x-y)G^{(0)} + (i\hat\partial - m)
\frac{\delta G^{(0)}}{\delta\eta(y,x)}
+ \frac{e^2}{i}
\int dx_1 D^c_{\mu\nu}(x-x_1)\gamma_\mu
\frac{\delta}{\delta\eta(y,x)}\tr \gamma_\nu\frac{\delta G^{(0)}}
{\delta\eta(x_1,x_1)}= 0
\label{G(0)}
\ee
and
 \be
\delta(x-y)G^{(0)} + (i\hat\partial - m)
\frac{\delta G^{(0)}}{\delta\eta(y,x)}
+ ie^2
\int dx_1 D^c_{\mu\nu}(x-x_1)\gamma_\mu
\frac{\delta}{\delta\eta(x_1,x)}\gamma_\nu\frac{\delta G^{(0)}}
{\delta\eta(y,x_1)}= 0.
\label{G(0)F}
\ee
 The functional
\be
G^{(0)}=\exp\Big\{\,\mbox{Tr}\,S^{(0)}\star\eta\Big\},
\label{G0F}
\ee
is a solution for  both these equations,
but, whereas the solution of the characteristic equation  corresponding
to eq.(\ref{G(0)}) is the
free propagator $S^{(0)}=S^c$ with condition
(\ref{condS}), a characteristic equation corresponding to eq.(\ref{G(0)F})
has the form
\be
[S^{(0)}]^{-1}(x)=(m-i\hat\partial)\delta(x)
-ie^2D^c_{\mu\nu}(x)\gamma_\mu S^{(0)}(x)\gamma_\nu,
\label{charF}
\ee
i.e., it is a non-trivial nonlinear equation for $S^{(0)}$,
and so we name this scheme the calculations over nonperturbative
vacuum as distinct from above scheme of eq.(\ref{SDE}) which is based
on the free solution  $S^c$.

A detailed discussion of eq.(\ref{charF}) is given below.
At the moment we note that at   $m=0$ (chiral limit) in
transverse gauge $d_l=0$ this equation possesses a simple solution
\be
 S^{(0)}=-1/i\hat\partial.
\label{Sc0} \ee Really, in coordinate space\footnote {The case
$n=2$ is discussed below, see Conclusion}       at $n>2$
\be
D^c_{\mu\nu}(x)=\frac{e^{-i\pi n/2}\Gamma(\frac{n}{2}-1)}
{4i\pi^{n/2}(x^2-i0)^{n/2-1}}\Big[\frac{1+d_l}{2}g_{\mu\nu}
+(1-d_l)\big(\frac{n}{2}-1\big)\frac{x_\mu x_\nu}{x^2-i0}\Big].
\label{Dcx}
\ee
In transverse gauge $d_l=0$ the function $D^c_{\mu\nu}(x)$
possesses an important property ("$\hat x$-transversality")
\be
D^c_{\mu\nu}(x)\gamma_\mu\hat x\gamma_\nu = 0.
\label{Dcxprop}
\ee
From this property the existence of solution (\ref{Sc0}) at $m=0$
follows immediately.

At $m\neq 0$  solving of the characteristic equation
is rather complicated problem which needs  application
of some approximate or numerical methods.
The search of solutions,
breaking the chiral symmetry, is also of great interest
(see below, Section 6).

An iteration equation for the nonperturbative vacuum in correspondence
with eqs.(\ref{SDEF}) and (\ref{G(0)F}) is
\ba
\delta(x-y)G^{(i)} + (i\hat\partial-m)\frac{\delta G^{(i)}}{\delta\eta(y,x)}
+ ie^2
\int dx_1 D^c_{\mu\nu}(x-x_1)\gamma_\mu
\frac{\delta}{\delta\eta(x_1,x)}\gamma_\nu\frac{\delta G^{(i)}}
{\delta\eta(y,x_1)}=
\nonumber\\
= \int dx_1 \Big\{\eta(x,x_1)
\frac{\delta G^{(i-1)}}{\delta\eta(y,x_1)} +
e D^c_{\mu\nu}(x-x_1)J_\nu(x_1)\gamma_\mu
\frac{\delta G^{(i-1)}}{\delta\eta(y,x)}\Big\}.
\label{G(i)F}
\ea
A solution of the first-step equation is
$G^{(1)}=P^{(1)}G^{(0)}$, where
\be
P^{(1)}= \frac{1}{2}\eta\star S^{(1)}_2\star\eta +
S^{(1)}\star\eta
+ J_\mu\star F_\mu^{(1)}\star\eta.
\label{P1F}
\ee
With taking into account  leading approximation
(\ref{G(0)F}) and characteristic equation
(\ref{charF}) we obtain for three-point function  $F_\lambda^{(1)}$,
two-electron function $S^{(1)}_2$ and propagator correction $S^{(1)}$
the following equations
\ba
F_\lambda^{(1)}(z; x,y)= -e\int dx_1 D^c_{\lambda\mu}(z-x_1)
S^{(0)}(x-x_1)\gamma_\mu S^{(0)}(x_1-y)+
\nonumber\\+
ie^2\int dx_1dy_1 D^c_{\mu\nu}(x_1-y_1)S^{(0)}(x-x_1)\gamma_\mu
F_\lambda^{(1)}(z; x_1,y_1)\gamma_\nu S^{(0)}(y_1-y),
\label{F(1)}
\ea
\ba
S_2^{(1)}(x,y;x',y')= - S^{(0)}(x-y') S^{(0)}(x'-y)+
\nonumber\\+
ie^2\int dx_1dy_1 D^c_{\mu\nu}(x_1-y_1)S^{(0)}(x-x_1)\gamma_\mu
S_2^{(1)}( x_1,y_1;x',y')\gamma_\nu S^{(0)}(y_1-y),
\label{S2(1)}
\ea
\ba
S^{(1)}(x-y)=
ie^2\int dx_1dy_1 D^c_{\mu\nu}(x_1-y_1)S^{(0)}(x-x_1)\gamma_\mu
S_2^{(1)}( x_1,y_1;y_1,y)\gamma_\nu+
\nonumber\\
+
ie^2\int dx_1dy_1 D^c_{\mu\nu}(x_1-y_1)S^{(0)}(x-x_1)\gamma_\mu
S^{(1)}(x_1-y_1)\gamma_\nu S^{(0)}(y_1-y).
\label{S(1)}
\ea

The first-step equations (\ref{F(1)})--(\ref{S(1)}) are much more
complicated in comparison with the first-step equations over
perturbative vacuum considered above. In diagram language\\
perturbative-vacuum equations (\ref{DF})--(\ref{SF}) correspond to
the summation of "chains", and its solutions can be easily written
out in a general form (see (\ref{F1})--(\ref{S1})). Equations
(\ref{F(1)})--(\ref{S(1)}) and characteristic equation
(\ref{charF}) in the diagram language correspond to the well-known
ladder approximation. Such type equations for separate Green
functions repeatedly were written out and investigated in
literature (see \cite{Nambu}--\cite{BJW}, and also \cite{Mir} and
references therein) as the simplest nonperturbative approximation
for exact Dyson equations \cite{Dyson} which constitute  an
infinite system of engaging equations. In our treatment, in
contrast to preceding investigations, these equation are not a
result of more or less arbitrary truncation of Dyson equations,
but are a consistent part of the iteration scheme. This
circumstance is very important for solving  such  problems as
taking into account  requirements of gauge invariance and
renormalizability, which often became stumbling-blocks at
investigation of nonperturbative approximations. So, for example,
one can often encounter the statement that equation (\ref{charF})
for electron propagator contradicts
 Ward identity  (\ref{WF}) and, consequently is not consistent
with gauge invariance (see discussion of this problem in \cite{Mir}).
However, one should not forget that a comparison of an {\it approximate}
equation (\ref{charF}) with the {\it exact} Ward identity (\ref{WF})
is not correct, and when an iteration scheme is absent the
formulation of this problem, strongly speaking, is incorrect itself.
In the framework of our iteration scheme this problem is solved very simply.
As  follows from eq.
(\ref{genWi}), Ward identity  $F_\mu^{(i)}$ in this framework has the form
\be
\frac{i}{d_l}\partial^2\partial_\nu F_\nu^{(i)}(x; x',y')
= e[\delta(x-y')-\delta(x-x')]S^{(i-1)}(x'-y').
\label{WF(i)}
\ee
It is easy to prove that for  $i=1$ relation (\ref{WF(i)}) and
equation  (\ref{charF}) are consistent with
the first-step equation (\ref{F(1)})
for $F_\mu^{(1)}$, i.e. the requirement of gauge invariance is fulfilled.

Turn to the renormalization. A renormalized equation for
$\delta G/\delta J_\mu$ has the form
\be
  \frac{1}{i}\frac{\delta G}{\delta J_\mu(x)}=
  -\int dx_1 D^c_{\mu\nu}(x-x_1\mid Z_3)\Big\{J_\nu(x_1)G+
  Z_1ie_r\tr \gamma_\nu\frac{\delta G}{\delta\eta(x_1,x_1)}\Big\}
  \label{dG/dJr}
  \ee
where
 $$
 D^c_{\mu\nu}( Z_3)=
[Z_3(g_{\mu\nu}\partial^2-\partial_\mu\partial_\nu)+
\frac{1}{d_l}\partial_\mu\partial_\nu]^{-1}.
$$
Correspondingly,  renormalized SDE
(\ref{SDEF}) is
 \ba
\delta(x-y)G + Z_1(i\hat\partial-m_r)\frac{\delta G}{\delta\eta(y,x)}
- \delta m\frac{\delta G}{\delta\eta(y,x)}+
\nonumber\\
+ i(e_rZ_1)^2\int dx_1 D^c_{\mu\nu}(x-x_1\mid Z_3)\gamma_\mu
\frac{\delta}{\delta\eta(x_1,x)}\gamma_\nu\frac{\delta G}
{\delta\eta(y,x_1)}=
\nonumber\\
= \int dx_1 \Big\{\eta(x,x_1)
\frac{\delta G}{\delta\eta(y,x_1)} +
Z_1e D^c_{\mu\nu}(x-x_1\mid Z_3)J_\nu(x_1)\gamma_\mu
\frac{\delta G}{\delta\eta(y,x)}\Big\}.
\label{SDEFr}
\ea
A renormalized leading-approximation equation is
 \ba
\delta(x-y)G^{(0)}+Z_1^{(0)}
(i\hat\partial-m_r)\frac{\delta G^{(0)}}{\delta\eta(y,x)}
- \delta m^{(0)}\frac{\delta G^{(0)}}{\delta\eta(y,x)}+
\nonumber\\
+ i(e_rZ_1^{(0)})^2\int dx_1 D^c_{\mu\nu}(x-x_1\mid Z_3^{(0)})\gamma_\mu
\frac{\delta}{\delta\eta(x_1,x)}\gamma_\nu\frac{\delta G^{(0)}}
{\delta\eta(y,x_1)}= 0.
\label{G(0)Fr}
\ea
Equation (\ref{G(0)Fr}) has a solution in the form of the  same functional
(\ref{G0F}), where $S^{(0)}$ is a solution of renormalized
characteristic equation
\be
[S^{(0)}]^{-1}(x)=
(Z_1^{(0)}(m_r-i\hat\partial)+\delta m^{(0)})\delta(x)
-i(e_rZ_1^{(0)})^2D^c_{\mu\nu}(x\mid Z_3^{(0)})
\gamma_\mu S^{(0)}(x)\gamma_\nu.
\label{charFr}
\ee
A renormalized photon propagator is defined by Dyson formula
\be
D_{\mu\nu}(x-y)
 = D^c_{\mu\nu}(x-y\mid Z_3)
 +ie_rZ_1\int dx_1 D^c_{\mu\rho}(x-x_1\mid Z_3)\tr
 \gamma_\rho F_\nu(y;x_1,x_1).
\label{Dr}
\ee
Since $F_\nu^{(0)}\equiv 0$ at the leading approximation, then
$$
D_{\mu\nu}^{(0)}
 = D^c_{\mu\nu}( Z_3^{(0)})=
 \frac{1}{Z_3^{(0)}\partial^2}(g_{\mu\nu}-\frac{\partial_\mu\partial_\nu}
 {\partial^2})+d_l\frac{\partial_\mu\partial_\nu}
 {(\partial^2)^2}.
$$
From the normalization condition for the photon propagator we obtain
$$
Z_3^{(0)}=1,
$$
therefore, in all above formulae one can replace
$D^c_{\mu\nu}( Z_3^{(0)})$ with $D^c_{\mu\nu}$.

In general case the renormalized photon propagator in the momentum space is
$$
D_{\mu\nu}(k) = D_{\mu\nu}^c(k\mid Z_3)+ie_rZ_1D_{\mu\rho}^c(k\mid Z_3)
\int\frac{dp}{(2\pi)^n}\tr\Big\{\gamma_\rho F_\nu(k;p)\Big\},
$$
where $F_\nu(k;p)$ is a Fourier-image of the three-pointer:
\be
F_\nu(z;x,y)=
\int\frac{dp}{(2\pi)^n}\frac{dq}{(2\pi)^n}\frac{dk}{(2\pi)^n}
e^{-ipx+iqy-ikz}(2\pi)^n\delta(p-q+k)
F_\nu(k;p).
\label{fourier}
\ee
 The transversality follows from Ward identities:
$$
ie_rZ_1\int\frac{dp}{(2\pi)^n}\tr\Big\{\gamma_\rho F_\nu(k;p)\Big\}=
\pi_{\rho\nu}f(k^2)
$$
and, taking into account given definition of function
$f(k^2)$, we obtain
\be
\label{D(k)}
D_{\mu\nu}(k)=-\frac{1+f(k^2)}{Z_3k^2}\pi_{\mu\nu}-d_l\frac{k_\mu k_\nu}
{(k^2)^2}.
\ee
  Zero mass normalization condition ("photon normalization ")
gives us the following
\be
Z_3=1+f(0).
\label{Z3}
\ee

The constant $Z_1$  and the mass-renormalization counterterm $\delta m$
are connected with the normalization of the electron propagator.
In general case the electron propagator in the momentum space is defined
by two scalar functions
$$
S(p)=\frac{1}{b(p^2)-a(p^2)\hat p}=\frac{b(p^2)+a(p^2)\hat p}
{b^2(p^2)-a^2(p^2)p^2}.
$$
If the propagator $S(p)$ possesses a pole at the point $p^2=m^2_r$, then
the normalization conditions in this point are
\be
b(m^2_r)=m_ra(m^2_r),
\label{normSa}
\ee
\be
a(m^2_r)+2m^2_ra'(m^2_r)-2m_rb'(m^2_r)=1
\label{normSb}.
\ee

A chiral limit for the renormalized theory means that the
chiral-non-symmetric terms disappear from the
renormalized Lagrangian (\ref{lagrangianr}):
 \be
Z_1m_r+\delta m\rightarrow 0.
\label{chirlim}
\ee
For  transverse gauge   $d_l=0$
renormalized characteristic equation (\ref{charFr})
 has a chiral-symmetric solution at the chiral limit
\be
S^{(0)}=-\frac{1}{Z_1^{(0)}\hat p}.
\label{Sc0r}
\ee
For this solution
normalization conditions  (\ref{normSa}) and (\ref{normSb}) give us
the relations
$$
Z_1^{(0)}=1,\;\;m_r=\delta m^{(0)}=0.
$$
Hence, this solution coincide with unrenormalized solution  (\ref{Sc0}).

Renormalized equations for the first-step coefficient functions are
\ba
F_\lambda^{(1)}(z; x,y)= -e_rZ_1^{(0)}\int dx_1 D^c_{\lambda\mu}(z-x_1)
S^{(0)}(x-x_1)\gamma_\mu S^{(0)}(x_1-y)+
\nonumber\\+
i(e_rZ_1^{(0)})^2
\int dx_1dy_1 D^c_{\mu\nu}(x_1-y_1)S^{(0)}(x-x_1)\gamma_\mu
F_\lambda^{(1)}(z; x_1,y_1)\gamma_\nu S^{(0)}(y_1-y),
\label{F(1)r}
\ea
\ba
S_2^{(1)}(x,y;x',y')= - S^{(0)}(x-y') S^{(0)}(x'-y)+
\nonumber\\+
i(e_rZ_1^{(0)})^2
\int dx_1dy_1 D^c_{\mu\nu}(x_1-y_1)S^{(0)}(x-x_1)\gamma_\mu
S_2^{(1)}( x_1,y_1;x',y')\gamma_\nu S^{(0)}(y_1-y),
\label{S2(1)r}
\ea
\ba
S^{(1)}(x-y)=
i(e_rZ_1^{(0)})^2
\int dx_1dy_1 D^c_{\mu\nu}(x_1-y_1)S^{(0)}(x-x_1)\gamma_\mu
S_2^{(1)}( x_1,y_1;y_1,y)\gamma_\nu+
\nonumber\\
+\int dx_1 S^{(0)}(x-x_1)\bigg\{
[Z_1^{(1)}(i\hat\partial-m_r)+\delta m^{(1)}]
S^{(0)}(x_1-y)+
\nonumber\\
+ie_r^2Z_1^{(0)}\int dy_1[2Z_1^{(1)}D_{\mu\nu}^c(x_1-y_1)+
Z_1^{(0)}D_{\mu\nu}^c(x_1-y_1\mid Z_3^{(1)})]
\gamma_\mu S^{(0)}(x_1-y_1)\gamma_\nu S^{(0)}(y_1-y)\bigg\}+
\nonumber\\+
i(e_rZ_1^{(0)})^2
\int dx_1dy_1 D^c_{\mu\nu}(x_1-y_1)S^{(0)}(x-x_1)\gamma_\mu
S^{(1)}(x_1-y_1)\gamma_\nu S^{(0)}(y_1-y).
\label{S(1)r}
\ea
As we see, the gauge invariance applies rather strict conditions:
constant $Z_1^{(0)}$ is the unique renormalization constant for two
first-step equations (\ref{F(1)r}) and (\ref{S2(1)r}).
In the equation for $S^{(1)}$ new renormalization constants appear:
there are
$Z_1^{(1)}$,  $Z_3^{(1)}$  (the last one is fixed by
condition (\ref{Z3})) and the first-step mass-renormalization counterterm
$\delta m^{(1)}$.

\section{Vertex}

Let us go from the three-point function $F_\lambda$ to the amputated
three-pointer --- vertex function (\ref{Gamma}).
In correspondence with eq.(\ref{F(1)r}) the first-step equation
for the vertex is
\ba
\Gamma_\lambda(z;x,y)=-e\gamma_\lambda\delta(x-y)\delta(x-z)+
\nonumber\\
+ie^2D^c_{\mu\nu}(x-y)\int dx_1dy_1\gamma_\mu S(x-x_1)
\Gamma_\lambda(z;x_1,y_1)S(y_1-y)\gamma_\nu.
\label{Gamma(1)}
\ea
In this Section we shall use simplified notations
$$
\Gamma_\lambda\equiv\Gamma_\lambda^{(1)},\;\;
e\equiv e_rZ_1^{(0)},\;\;
S\equiv S^{(0)}.
$$
In the momentum space equation (\ref{Gamma(1)}) has the form
 \be
\Gamma_\lambda(k;p)=-e\gamma_\lambda
+ie^2\int \frac{dp'}{(2\pi)^n}  D^c_{\mu\nu}(p-p') \gamma_\mu S(p')
\Gamma_\lambda(k;p')S(p'+k)\gamma_\nu.
\label{Gamma(1)f}
\ee
Here $k$ is the photon momentum, $p$ is the electron momentum
(see definition (\ref{fourier}) of the Fourier-image of three-pointer).
Equation (\ref{Gamma(1)f}) is known in literature as Edwards equation
\cite{Edwards}. Note again that in our approach this equation is
not a result of arbitrary truncation, but is a consequence of the iteration
scheme equations. Electron propagator $S$ in this equation is a solution
of characteristic equation (\ref{charFr}) which defines the leading
vacuum approximation.

We consider a solution of vertex equation
(\ref{Gamma(1)}) at small $k$ with above mentioned chiral-symmetric
propagator (\ref{Sc0r}) in the transverse gauge  $d_l=0$.
To solve this equation it is convenient to introduce a function
$\Phi_\lambda(k;p)$ which is defined by the relation
$$
\Gamma_\lambda(k;p)\equiv \hat p\Phi_\lambda(k;p)(\hat p+\hat k).
$$
Expand $\Phi_\lambda$ near $k=0$
$$
\Phi_\lambda(k;p)=\Phi_\lambda(p)+k_\rho\Phi_{\lambda\rho}(p)
+\cdots
$$
Here
$$
\Phi_\lambda(p)\equiv\Phi_\lambda(0;p),\;\;
\Phi_{\lambda\rho}(p)\equiv\partial^k_\rho\Phi_\lambda(k;p)\vert_{k=0}.
$$
At that
\be
\Gamma_{\lambda\rho}(p)\equiv\partial^k_\rho\Gamma_\lambda(k;p)\vert_{k=0}
=\hat p\Phi_{\lambda\rho}(p)\hat p+\hat p\Phi_\lambda(p)\gamma_\rho.
\label{Gammalr}
\ee
Equations for
$\Phi_\lambda(p)$ and $\Phi_{\lambda\rho}$ are
\be
\hat p\Phi_\lambda(p)\hat p=-e\gamma_\lambda+
ie^2_r\int \frac{dp'}{(2\pi)^n}  D^c_{\mu\nu}(p-p') \gamma_\mu
\Phi_\lambda(p')\gamma_\nu
\label{Phi0}
\ee
and
\be
\hat p\Phi_{\lambda\rho}(p)\hat p=
-\hat p\Phi_\lambda(p)\gamma_\rho+
ie^2_r\int \frac{dp'}{(2\pi)^n}  D^c_{\mu\nu}(p-p') \gamma_\mu
\Phi_{\lambda\rho}(p')\gamma_\nu.
\label{Phi1}
\ee

It is more convenient to solve eq.(\ref{Phi0}) in the coordinate space:
\be
-\hat\partial\Phi_\lambda(x)\hat\partial=-e\gamma_\lambda\delta(x)+
ie^2_r D^c_{\mu\nu}(x) \gamma_\mu
\Phi_\lambda(x)\gamma_\nu.
\label{Phi0x}
\ee
Here $D^c_{\mu\nu}(x)$ is defined by eq.(\ref{Dcx}).
Due to the property of
"$\hat x$-transversality" (see (\ref{Dcxprop})) all iterations of
eq.(\ref{Phi0x}) turn into zero. Indeed, a zero approximation is
$$
\Phi^0_\lambda(p)=-e\frac{1}{\hat p}\gamma_\lambda\frac{1}{\hat p}=
e\partial^p_\lambda\frac{1}{\hat p}
$$
and in $x$-space
$$
\Phi^0_\lambda(x)=e\frac{ie^{-i\pi n/2}\Gamma(n/2)}
{2\pi^{n/2}}\frac{x_\lambda\hat x}{(x^2-i0)^{n/2}}\sim \hat x.
$$
Consequently
$$
\Phi_\lambda=\Phi_\lambda^0
$$
and
\be
\Gamma_\lambda(p)=-e\gamma_\lambda.
\label{Gamma0p}
\ee

Let us go to  solving  eq.(\ref{Phi1}) for
$\Phi_{\lambda\rho}$. (We consider now the four-dimensional case $n=4$.)
This equation is also more convenient for solving in the coordinate
space, where it has the form
\be
-\hat\partial\Phi_{\lambda\rho}(x)\hat\partial=
\frac{e}{2\pi^2}\frac{\gamma_\lambda\hat x\gamma_\rho}
{(x^2-i0)^2}+
ie^2_r D^c_{\mu\nu}(x) \gamma_\mu
\Phi_{\lambda\rho}(x)\gamma_\nu.
\label{Phi1x}
\ee
(We have taken into account the result of solving equation for
$\Phi_\lambda(p)$.)
Here
\be
D^c_{\mu\nu}(x)= \frac{1}{4i\pi^2(x^2-i0)}
\big(\frac{g_{\mu\nu}}{2}+\frac{x_\mu x_\nu}{x^2-i0}\big).
\label{Dc4}
\ee
Expanding $\Phi_{\lambda\rho}$ over spinor structures
$$
\Phi_{\lambda\rho}=\Phi_{\lambda\rho\sigma}\gamma_\sigma
+\Phi_{\lambda\rho\sigma}^5\gamma_5\gamma_\sigma
$$
we obtain two equations
\ba
(g_{\sigma\tau}\partial^2-2\partial_\sigma\partial_\tau)
\Phi_{\lambda\rho\sigma}(x)=
\frac{e}{2\pi^2}\frac{g_{\lambda\tau}x_\rho-g_{\lambda\rho}x_\tau
+g_{\rho\tau}x_\lambda}{(x^2-i0)^2}-
\nonumber\\
-\frac{2\alpha_r}{\pi}\frac{1}{x^2-i0}\big(g_{\sigma\tau}-
\frac{x_\sigma x_\tau}{x^2-i0}\big)\Phi_{\lambda\rho\sigma}(x),
\label{PhiS}
\ea
\be
(g_{\sigma\tau}\partial^2-2\partial_\sigma\partial_\tau)
\Phi_{\lambda\rho\sigma}^5(x)=
\frac{ie}{2\pi^2}\epsilon_{\lambda\rho\sigma\tau}
\frac{x_\sigma}{(x^2-i0)^2}
-\frac{2\alpha_r}{\pi}\frac{1}{x^2-i0}\big(g_{\sigma\tau}-
\frac{x_\sigma x_\tau}{x^2-i0}\big)\Phi_{\lambda\rho\sigma}^5(x).
\label{Phi5}
\ee
Here
$$
\alpha_r=\frac{e^2_r}{4\pi}.
$$

As for equation for $\Phi_\lambda$  iterations of eq.
(\ref{PhiS}) for $\Phi_{\lambda\rho\sigma}$ turn into zero,
and its solution is
$$
\Phi_{\lambda\rho\sigma}(x)=-\frac{e}{4\pi^2}
\frac{x_\lambda x_\rho x_\sigma}{(x^2-i0)^2}.
$$
Equation (\ref{Phi5}) for  $\Phi_{\lambda\rho\sigma}^5$ possesses
a solution
$$
\Phi_{\lambda\rho\sigma}^5(x)=
\frac{ie}{4\pi(2\pi-\alpha_r)}\epsilon_{\lambda\rho\sigma\tau}
\frac{x_\tau}{x^2-i0}.
$$
Going to the momentum space we finally obtain for
$\Gamma_{\lambda\rho}$ (see (\ref{Gammalr})):
\be
\Gamma_{\lambda\rho}(p)=
ie\frac{\alpha_r}{2\pi-\alpha_r}
\epsilon_{\lambda\rho\sigma\tau}\gamma_5\gamma_\sigma\frac{p_\tau}{p^2+i0}.
\label{DGamma}
\ee

 Matrix element of vertex function
$\Gamma_\lambda(k;p)$ at small $k_\mu$ defines two formfactors
$f_1$ and $f_2$
 \be
 \bar u(q)\Gamma_\lambda u(p)\simeq-e_rf_1\bar u(q)\gamma_\lambda u(p)
 -\frac{e_r}{2m_r}f_2\bar u(q)[\gamma_\lambda,\hat k] u(p),
 \label{forms}
 \ee
 which connected correspondingly with a physical charge and a
 magnetic moment (see, for example,
 \cite{BogSh}). Here $u(p)$ is a solution of Dirac equation
 $(\hat p-m_r)u(p)=0$. Comparing with calculated vertex we obtain
 at  $k=0$
 $$
 e_rf_1=e=e_rZ_1^{(0)}.
 $$
 For the  normalization on the charge
 $e_r$ we put $f_1=1$ and, consequently,
 $$
 Z_1^{(0)}=1.
 $$
This very important fact means that the normalization of the vertex
function on the renormalized charge gives in our approach the same
value of renormalization constant $Z_1$ as
wave-function renormalization condition (\ref{normSb}),
i.e. our calculational scheme is consistent with requirements
which imposed by gauge invariance.

 As for the second formfactor   $f_2$, which defines a correction
 to the magnetic moment, note, that the massless particle, of course,
 has no magnetic moment. However, the calculated first term of
  vertex function expansion in $k_\mu$ (see (\ref{DGamma}))
 permits making a nonperturbative estimate for anomalous magnetic
 moment in chiral limit.

 Let us define the problem more exactly. In expansion of $f_2$ in
 orders of  $\alpha_r$
  $$
 f_2=C_1\frac{\alpha_r}{\pi}+C_2\Big(\frac{\alpha_r}{\pi}\Big)^2+\cdots
 $$
 coefficients $C_i$, starting with the second one, are mass-depended.
(So, they have even different signs for muon and electron, see, for example,
 \cite{AkhBe}.) Nevertheless, one can, in principle,
 to consider a peculiar "chiral limit"
 for these coefficient, i.e. their values at $m_r\rightarrow 0$.
 It is this estimate which we are talking about.
 To obtain this estimate let us make
 an identical transformation of the second term in eq.(\ref{forms})
 \be
 \frac{1}{m_r}\bar u(q)[\gamma_\lambda,\hat k] u(p)=
 \frac{1}{p^2}\bar u(q)[\gamma_\lambda,\hat k]\hat p u(p).
 \label{identity}
 \ee
 For massive particles it is an identity due to Dirac equation
 for $u(p)$. However, for r.h.s. of eq.(\ref{identity}) one can use
 result (\ref{DGamma}) which was obtained for massless particle.
 Comparing coefficients at $\epsilon_{\lambda\rho\sigma\tau}$,
 we obtain
  \be
 f_2=\frac{\alpha_r}{2\pi-\alpha_r}.
 \label{f2}
 \ee
The first term of the expansion in $\alpha_r$
 coincides with the Schwinger correction
 (see, for example, \cite{AkhBe}).
 At formal limit $\alpha_r\rightarrow\infty$ the full magnetic moment
 in correspondence with eq.(\ref{f2}) turn to zero. It is interesting
 to note, that similar screening of chromomagnetic moments of quarks
 was pointed out in some models of nonperturbative quantum chromodynamics
 \cite{AQCD} and also in relativistic quarkonium model  \cite{Fau}.

 \section{Dynamical chiral symmetry breaking}

In this Section we consider a problem of dynamical chiral symmetry
breaking (DCSB) for QED in the framework of proposed calculational scheme
over the nonperturbative vacuum. Note, that our present consideration
 does not cover all aspects of this complicated
problem, in particular, we do not consider a question of connection
of DCSB with triviality problem. A consideration of this and
a number of other questions imposes an investigation of the
first-step equations for higher Green functions. We limit ourselves
to investigation of the leading-approximation equation.

A renormalized equation of leading approximation for the electron propagator
has the form
\be
S^{-1}(x)=(m-Zi\hat\partial)\delta(x)
-ie^2D^c_{\mu\nu}(x)\gamma_\mu S(x)\gamma_\nu.
\label{charr}
\ee
Here and everywhere in this Section we denote
$$
S\equiv S^{(0)},\;\;Z\equiv Z^{(0)}_1,\;\;
m\equiv Z^{(0)}_1m_r+\delta m^{(0)},\;\;
e\equiv e_rZ^{(0)}_1,
$$
and the free propagator  $D^c_{\mu\nu}(x)$ is defined by eq.(\ref{Dcx}).

In general case (if one does not consider parity breaking solutions)
$S$ can be represented as
$$
S=i\hat\partial A+B,
$$
where $A$ and  $B$ are scalar functions of unique variable.
Correspondingly, for the inverse propagator we have
$$
S^{-1}=-i\hat\partial a+b,
$$
where in the momentum space the functions
$a$ and $b$ are connected with the functions
$A$ and  $B$ by relations
\be
A=\frac{a}{b^2-a^2p^2},\;\; B=\frac{b}{b^2-a^2p^2}.
\label{AB}
\ee
Taking into account
eq.(\ref{Dcx}) and properties of Dirac matrices,
characteristic equation (\ref{charr}) in
the coordinate space can be represented
as a system of two equations
\be
i\hat\partial a=Zi\hat\partial\delta(x)-
2\alpha d_l\frac{e^{-i\pi n/2}\Gamma(\frac{n}{2})}
{[\pi(x^2-i0)]^{n/2-1}}\cdot i\hat\partial A,
\label{eqa}
\ee
\be
b= m\delta(x)+
\alpha(1-n-d_l)\frac{e^{-i\pi n/2}\Gamma(\frac{n}{2}-1)}
{[\pi(x^2-i0)]^{n/2-1}}\cdot B.
\label{eqb}
\ee
Here
$$
\alpha\equiv\frac{e^2}{4\pi}.
$$

 As is evident from eqs. (\ref{eqa})-(\ref{eqb}),  two special gauges exist
for  characteristic equation (\ref{charr})

1)the transverse Landau gauge $d_l=0$.
In this gauge eq.(\ref{eqa}) becomes trivial and possesses
a solution $a=Z\delta(x).$

2)the gauge $d_l=1-n$. In this gauge\footnote{At $n=4$ it is
Solovev-Yennie gauge \cite{Sol}.}
  $b=m\delta(x)$
{\it in case if the product $B(x)\cdot (x^2-i0)^{1-n/2}$ is well-defined
 in a sense of generalized functions,
i.e., divergences are absent}. Otherwise, in contrast with the Landau
gauge, we have an uncertainty of the type
$0\cdot\infty$. Its evaluation is performed by a regularization,
and we can only  say that in this gauge $b(p^2)=\mbox{const}$.

Below in this Section we shall consider the four-dimensional case.
Multiplying eqs.(\ref{eqa}) and (\ref{eqb}) by $x^2$,
we obtain the equations
\be
x^2i\hat\partial a=-\frac{2\alpha}{\pi}d_li\hat\partial A,
\label{eqa4}
\ee
\be
x^2b=-\frac{\alpha}{\pi}(3+d_l)B.
\label{eqb4}
\ee
Such a multiplication can be considered as some regularization
of potentially singular products in the r.h.s. of eqs.
(\ref{eqa}) and (\ref{eqb}). More, going to  $p$-space and
taking into account eq.(\ref{AB}), we obtain for
$a(p^2)$ and $b(p^2)$
a system of ordinary differential equations
\be
t\frac{d^2a}{dt^2}+3\frac{da}{dt}=\frac{\alpha}{2\pi}d_l
\frac{a}{b^2-a^2t},
\label{eqadiff}
\ee
\be
t\frac{d^2b}{dt^2}+2\frac{db}{dt}=\frac{\alpha}{4\pi}(3+d_l)
\frac{b}{b^2-a^2t}.
\label{eqbdiff}
\ee
Here $t=p^2$.

This very complicated system of nonlinear  differential equations
has the simplest form in the transverse Landau gauge $d_l=0$.
Below we shall exploit this gauge. As  was mentioned earlier,
in this gauge $a=Z$, and system
(\ref{eqadiff})-(\ref{eqbdiff}) is  reduced to the equation for $b$
\be
t\frac{d^2b}{dt^2}+2\frac{db}{dt}=\frac{3\alpha}{4\pi}
\frac{b}{b^2-Z^2t}.
\label{eqb0}
\ee
This equation always has the trivial solution $b\equiv0$,
which corresponds to above
mentioned chiral-symmetric solution (\ref{Sc0r}).
The existence of non-trivial solutions is also possible,
and their asymptotics is not difficult to find.
Really, at $t\rightarrow\infty$  two variants are possible:

a) $b^2\sim t$;

b) $b^2\ll\mid t\mid$.

It is not difficult to prove that the third possibility
$b^2\gg\mid t\mid$ up to logarithms leads to the variant a).

Consider the variant b). Then in the ultraviolet region
$t\rightarrow\infty$ equation (\ref{eqb0}) is reduced to Euler equation,
and an asymptotic behavior of
$b(t)$ is
\be
b\sim t^{-1/2(1-\sqrt{1-3\alpha_r/\pi})}\;\;
\mbox{at} \;\; \alpha_r<\pi/3,
\label{as<}
\ee
\be
b\sim t^{-1/2}\log t\;\;\mbox{at} \;\; \alpha_r=\pi/3
\label{as=}
\ee
and
\be
b\sim t^{-1/2}\sin\big\{\frac{\sqrt{3\alpha_r/\pi-1}}{2}
\log t\big\}\;\;\mbox{at} \;\; \alpha_r>\pi/3.
\label{as>}
\ee
Note, that here
$\alpha_r\equiv \frac{e^2_r}{4\pi}$
is the renormalized fine structure constant.

 The existence of a critical value
\be
\alpha_r=\alpha_c\equiv\pi/3
\label{alfac}
\ee
should be  pointed out,
at which a change of ultraviolet behavior regime takes place.
The existence of this critical point was firstly pointed in work
\cite{Maskawa}. In works \cite{Fomin} such regime transition
was connected with DCSB in QED. (See also \cite{Mir},\cite{Rob}
where this approach is discussed in detail  and the extensive bibliography
is given. For latest developments of this approach see  \cite{Gus}.)
In all cited works a scheme with ultraviolet cutoff in Euclidean space
was exploited. A main calculational ansatz was a linearization of eq.
(\ref{eqb0}), which  consists in an approximation
\footnote{Note, that in the cutoff scheme  $Z=1$.}
$$
\frac{b}{b^2-t}\approx \frac{b}{m^2-t},
$$
in the r.h.s. of eq.(\ref{eqb0}). Here $m\equiv b(0)$.
After such a linearization the equation for $b(t)$ becomes
a hypergeometrical equation. Boundary conditions for this
equation are defined from the integral equation which is eq.(\ref{charF})
in the momentum space
\footnote{An essential point at the formulation of
these boundary conditions is exploiting of Euclidean version of
the theory.}.
An asymptotic behavior of solutions of linearized equation is
given by the same formulae (\ref{as<})-(\ref{as>}), and this fact
is the main argument in favor of the linearized version (at least
in the ultraviolet region). These solutions, of course, essentially
 depend on a cutoff parameter which is included in the boundary
 conditions. Nevertheless, one was succeeded in demonstrating
 the existence of a phase transition at the critical point $\alpha_c=\pi/3$.
This transition corresponds to DCSB. In the chiral limit
under the critical point
(weak coupling) only the chiral-symmetric solution exists, but
beyond the critical point (strong coupling), when the ultraviolet
asymptotics becomes oscillating, a solution with dynamical
mass $b\neq 0$ arises, i.e., the spontaneous breaking of chiral
symmetry takes a place.

We consider this problem in the pseudoeuclidean Minkowski space
for renormalized equation (\ref{charr}).\footnote{This point seems
to be essential since detailed investigations of unrenormalized
equation (\ref{charF}) in Euclidean space at  $m=0$ (see, for
example, \cite{Filip}) demonstrate the existence of non-pole
complex singularity for its solutions, and, hence, the Euclidean
rotation seems to be a problem.} At that we shall use another
approach to the investigation of the propagator equation, which,
nevertheless, is similar to the approach of above mentioned works
and also based on a linearization of nonlinear equation
(\ref{charr}). Since we work with renormalized theory, our results
are cutoff-independent. In general they are consistent with the
results of linearized unrenormalized theory in Euclidean space up
to the unique exception: in pre-critical region
$\alpha_r<\alpha_c$ DCSB is also possible, but under some
condition on value of $\alpha_r$. An investigation of this
condition requires  studying of equation for the three-point
function with the nonperturbative electron propagator and goes out
of the framework of this work.

We shall solve equation (\ref{charr}) in transverse gauge
 $d_l=0$ at $m\neq 0$ by iterations and a foundation for iterative
 solution will be an exact solution at   $m=0$

$$
S_0^{-1}=-Z\hat p.
$$
At that the linearization consists in rather natural procedure from
the point of view of the iterative solution: starting from the
representation of the inverse propagator
$$
S^{-1}=S_0^{-1}+\Sigma
$$
we approximate
$$
S=[S_0^{-1}+\Sigma]^{-1}\approx S_0-S_0\star\Sigma\star S_0=
-\frac{\hat p}{Z(p^2+i0)}-\frac{\hat p\Sigma\hat p}{Z^2(p^2+i0)^2}.
$$
It is clear that such an approximation can be
fully valid only in the ultraviolet region
 $\mid~p^2\mid\gg\mu^2$, where $\mu^2$ is some scale
 which can be treated as an infrared cutoff.
Since a nonperturbative region of QED is the ultraviolet region,
the natural supposition is in the
following: the ultraviolet behavior is crucial
 for basic nonperturbative effects including DCSB.

Using the property of $\hat x$-transversality (see (\ref{Dcxprop}))
it is easy to prove $\Sigma=1\cdot\sigma,$
and for scalar function $\sigma$ in the coordinate space we obtain
the equation
\be
\sigma(x^2)=m\delta(x)+\frac{3\alpha_r}{\pi}
\frac{1}{x^2-i0}\cdot\phi(x^2),
\label{sigma}
\ee
where
$$
\phi(p^2)\equiv\frac{\sigma(p^2)}{p^2+i0}.
$$

 The four-dimensional delta-function  $\delta(x)$ in the Minkowski space
 is a Lorentz-invariant distribution (see, for example,
 \cite{BLOT}), and to solve eq.(\ref{sigma}) it is convenient
 to use a representation of the delta function as a limit of
 a sequence of functions in $x^2$-variable. As such a representation
 we choose the formula
\be
\delta^4(x)=\frac{i}{\pi^2}\lim_{\lambda\rightarrow 0}
\lambda(x^2-i0)^{\lambda-2}
\label{delta}
\ee
which can be easy proved with the Fourier-transform of function
$(x^2-i0)^{\lambda-2}$ (see, for example,  \cite{Brych}).

A partial solution (a solution of inhomogeneous equation) will be
searched as
$$
\sigma_0(x^2)=C_0(x^2-i0)^{\lambda-2}
$$
at small $\lambda$. As we shall see below,  $1/\lambda$
plays the role of a regularization parameter.
Making the corresponding Fourier-transform (see \cite{Brych}),
we obtain
\be
\phi_0(x^2)=\frac{C_0}{4\lambda(1-\lambda)}(x^2-i0)^{\lambda-1},
\label{fi}
\ee
and, substituting into eq.
(\ref{sigma}), we finally obtain
$$
C_0=-\frac{4i m}{3\pi\alpha_r}\lambda^2+
{\cal O}(\lambda^3).
$$
Consequently, the solution of the inhomogeneous equation is
\be
\sigma_0(p^2)=-\frac{4\pi m\lambda }{3\alpha_r}.
\label{sigma0}
\ee

The full solution of eq.(\ref{sigma}) is a sum of $\sigma_0$ and
a general solution of homogeneous equation. A solution of
homogeneous equation is sought for in the form
$$
\bar\sigma(x^2)=C(x^2-i0)^{\beta-2}.
$$
Hence $\bar\phi(x^2)$ is defined by the same formula
(\ref{fi}) with the substitution $\lambda\rightarrow\beta$.
For parameter $\beta$ we obtain the equation
\be
\beta(1-\beta)=\frac{3\alpha_r}{4\pi}.
\label{beta}
\ee

If one chooses a solution with  $C=0$, i.e., limits oneself to
consideration of partial solution
(\ref{sigma0}), then normalization conditions (\ref{normSa}) and
(\ref{normSb}) give
\footnote{The formula for counterterm $\delta m$
demonstrates a role of
$\lambda$ as a parameter of regularization which removes
ultraviolet divergence. Therefore, the representation of delta-function
by formula
(\ref{delta}) can be considered as a special analytical regularization}
$$
\delta m^{(0)}=\frac{3\alpha_r}{4\pi\lambda}m_r,
\;\;Z=1$$
and, correspondingly, in this case
$$
S^{-1}= -\hat p+m_r.
$$
In the chiral limit (see (\ref{chirlim})), when $m\rightarrow 0$,
we obtain $m_r=0$, therefore this solution is a chiral-symmetric one.

But if one permits solutions with  $C\neq 0$, the situation is
essentially changed. Note at once, that the solutions of the homogeneous
equation in $p$-space are singular at $p=0$, but, due to the essence
of the approximation made, all subsequent formulae can be interpreted,
as was pointed out above, only at
$\mid p^2\mid\gg \mu^2$, where $\mu^2$ is an infrared cutoff.

So, in dependence of value of parameter $\alpha_r$, we have for electron
mass function:

1) At $\alpha_r<\alpha_c$ (weak coupling)
\be
b(p^2)=-\frac{4\pi m\lambda }{3\alpha_r}+C(p^2+i0)^{-\beta},
\label{b<}
\ee
where $\beta=\frac{1}{2}(1-\sqrt{1-\alpha_r/\alpha_c})$.
\footnote{We take a root of eq.(\ref{beta}) which corresponds to
the ultraviolet asymptotics of  exact solution, see
(\ref{as<}).}

2) At $\alpha_r=\alpha_c=\pi/3$
\be
b(p^2)=-\frac{4\pi m\lambda }{3\alpha_r}+
\frac{C\log\frac{p^2+i0}{M^2}}{(p^2+i0)^{1/2}}.
\label{b=}
\ee

3) At $\alpha_r>\alpha_c$ (strong coupling)
\be
b(p^2)=-\frac{4\pi m\lambda }{3\alpha_r}+
\frac{C\sin\big\{\frac{\omega}{2}\log\frac{p^2+i0}{M^2}\big\}}
{(p^2+i0)^{1/2}}.
\label{b>}
\ee
Here
$$\omega=\sqrt{\frac{\alpha_r}{\alpha_c}-1}.$$
$a=Z$ for any of three cases.

Below we consider the chiral limit
$m\equiv Zm_r+\delta m^{(0)}=0$.
Solution parameters are fixed by normalization conditions
(\ref{normSa})-(\ref{normSb}) of the electron propagator.
If (as in ultraviolet-cutoff scheme \cite{Fomin},\cite{Mir},
\cite{Gus}) one takes
$Z=1$, then the normalization conditions become
\be
b(m^2_r)=m_r, \;\; m_rb'(m^2_r)=0.
\label{norm1}
\ee
In the weak coupling case $(\alpha_r<\alpha_c)$
it  follows from the normalization conditions that either   $C=0$
or $m_r=0$, i.e. in this case DCSB-solutions   are absent.
In the critical case  $\alpha_r=\alpha_c$ a solution with $m_r\neq 0$
is possible which has the form
\be
b(p^2)=\frac{m^2_r}{(p^2+i0)^{1/2}}\Big(1+
\frac{1}{2}\log\frac{p^2+i0}{m_r^2}
\Big).
\label{b1=}
\ee
At $\alpha_r>\alpha_c$ (strong coupling) a DCSB-solution also exists:
\be
b(p^2)= \frac{m^2_r}{(p^2+i0)^{1/2}}
\Big[\cos\Big\{\frac{\omega}{2}\log\frac{p^2+i0}{m_r^2}\Big\}
+\frac{1}{\omega}\sin\Big\{\frac{\omega}{2}\log\frac{p^2+i0}{m_r^2}\Big\}
\Big].
\label{b1>}
\ee

As we can see, these results correspond to those of
the unrenormalized theory with a cutoff in the Euclidean space
(see   \cite{Fomin},\cite{Mir},\cite{Gus}).
However for the renormalized theory we have no any prior reason
to set $Z=1$. If one refuses  this condition, the results in
the weak coupling region can be essentially different.
Really, at $Z\neq 1$ the normalization conditions are
 \be
 b(m^2_r)=m_rZ, \;\; 2m_rb'(m^2_r)=Z-1.
\label{norm2}
\ee
At $\alpha_r<\alpha_c$, in contrast with the case  $Z=1$,
the existence of DCSB-solution  becomes possible. It has the form
\be
a=Z=\frac{1}{1+2\beta},\;\;b=Zm_r\Big(\frac{m^2_r}{p^2+i0}\Big)^\beta.
\label{S<}
\ee
At $\alpha_r=\alpha_c$ a solution is
\be
b(p^2)=\frac{m^2_r}{(p^2+i0)^{1/2}}\Big(Z+\big(Z-\frac{1}{2}\big)
\log\frac{p^2+i0}{m_r^2}
\Big).
\label{S=}
\ee
and, at last, at $\alpha_r>\alpha_c$
\be
b(p^2)= \frac{m^2_r}{(p^2+i0)^{1/2}}
\Big[Z\cos\Big\{\frac{\omega}{2}\log\frac{p^2+i0}{m_r^2}\Big\}
+\frac{1}{\omega}\big(2Z-1\big)
\sin\Big\{\frac{\omega}{2}\log\frac{p^2+i0}{m_r^2}\Big\}
\Big].
\label{S>}
\ee
($a=Z$ for two last cases, $Z$ is arbitrary.)

Therefore, with taking into account the renormalization freedom
DCSB  becomes possible in the weak coupling
region which corresponds to physical sector of the theory.
At that, however, an important limitation connected with
the gauge invariance arises.  The necessity for such a limitation
follows also from a consideration of switch-off-interaction limit.
Indeed, it follows from eq.(\ref{S<}) that at
$\alpha_r\rightarrow 0\;\;\;b\rightarrow m_r+{\cal  O}(\alpha_r)$,
i.e. either $m_r=0$, or DCSB does not disappear in the
switch-off-interaction limit. The last possibility looks quite
strange. A way out of this situation can be in the
following: all the picture
in the weak coupling region is realized only for some finite values of
$\alpha_r$.
A condition for such values is a normalization of the vertex
on physical charge $e_r$:
 \be
 \Gamma_\mu(k=0;p^2=m^2_r\mid \alpha_r, Z)=-e_r\gamma_\mu.
 \label{normGamma}
 \ee
Due to the gauge invariance the renormalization constant $Z$ is the same
as for the electron propagator equation, and substituting
the value   $Z=Z(\alpha_r)$
from eq.(\ref{S<}) we shall obtain an equation
for $\alpha_r$. Solutions of this equation will define "permitted"
values of the charge. A realization of this program requires
solving an equation for the vertex function with propagator (\ref{S<}).

At $\alpha_r\ge\alpha_á$ the situation  changes in principle:
in this case  $Z$ (at the leading approximation) is arbitrary,
and the normalization on the physical charge  fixes $Z$,
but does not lead to additional limitations on values of $\alpha_r$.

 \section*{Conclusion}

 Above results give us a reasoning to consider the calculations
 over the nonperturbative vacuum as an quite adequate scheme for
 calculations of nonperturbative effects in QED.
 A formulation of system of equations for Green functions at any
 step of this calculational scheme is technically as simple as
  the coupling constant perturbation theory
\footnote{Note, the coupling constant perturbation theory can be
considered as a partial case of the general iteration scheme based
on principles of Section 4. To obtain a perturbation theory series
it is enough consider the same iteration scheme with single
Grassman fermion sources over perturbative vacuum which
corresponds to leading approximation $G^{(0)}=1$  (see also
\cite{Ro1}-\cite{Ro2}).} and does not require anything except of
the "stupid differentiation". The equations themselves, of course,
are much more complicated, but, as we can see from the calculation
of the vertex function for chiral-symmetric solution (Section 5),
they possess much more simple solutions in comparison with
analogous equations for Green functions in nonperturbative
approximations which based on a more or less arbitrary truncation
of Dyson equations.

As regards the calculations of Section 6, which concern  the
DCSB problem, the one of main questions concerning the linearized
equation for electron propagator is a problem of a gauge dependence
of the results, in particular, of a gauge dependence of the critical
constant $\alpha_c$ (see \cite{Mir} and references therein).
In our treatment this equation is not a result of an arbitrary truncation
of Dyson equations but a consistent part of the iteration scheme,
and this question becomes unavoidable. In this connection we note
 that the linearization seems to be substantiate in Landau gauge only,
 and for other gauges the nonlinear effects scarcely can be
 taken into account with such simple ansatz.
As  has been mentioned above, the results of Section 6 in strong
coupling region coincide in essence to the results of so-called
rainbow approximation, or "quenched QED with bare vertex" (see
\cite{Mir}, \cite{Gus}). It is no wonder since the basic equation
for electron propagator is the same, and a difference consists in
the regularization and renormalization scheme. Generally, quenched
QED is an approximation of QED  with an effect of pair creation
being neglected. For the electron propagator this approximation
permits more general form of the vertex function. For example, the
use of Curtis-Pennington vertex \cite{CP} solves the problem of
gauge dependence of critical coupling $\alpha_c$, though is not
solve all the problems in strong coupling region (see
\cite{Williams} and references therein). Our approach is not an
approximation of QED but an iteration scheme, so we have not a
possibility for such generalization of leading order equation for
electron propagator. As a general remark, note, that the
strong-coupled four-dimensional QED is of pure theoretical
interest as a model gauge theory for investigation of chiral
symmetry breaking phenomenon. The results of various approaches
(such as worldline variational approach, high orders of
perturbation theory, quenched approximation etc.) essentially
differ in strong coupling region (see \cite{Schreiber} for
review), which indicates our present poor understanding of this
problem.

 In closing we shortly discuss a problem of DCSB for quantum
 chromodynamics (QCD). In contrast to QED, where nonperturbative
 effects (and DCSB among them) are defined by ultraviolet region,
 for asymptotically-free QCD a nonperturbative region is
 the low-energy infrared region. At that a mechanism of DCSB
 in QCD can be even more simple in comparison with QED, since
  effective taking into account of gluon self-action in
 the infrared region inevitably leads  to appearance of new
 dimensional parameters.

 As a simple example of a realization of similar DCSB mechanism
 by infrared singularities consider two-dimensional QED.
   The analogy among the nonperturbative effects
 of two-dimensional QED and those of  four-dimensional QCD
 was repeatedly pointed out. In spite of
  certainly limited nature of such analogy, an investigation
  of the nonperturbative dynamics on example of
  much more technically simple  QED can be useful for understanding
  of QCD in the nonperturbative region.

At $n=2$ the free propagator $D^c_{\mu\nu}(x)$ (see (\ref{Dcx}))
entering  equation
(\ref{charr}) in general case is infrared-singular, and it
is necessary to introduce some infrared cutoff.
 However, a gauge exists for which such a cutoff is unnecessary.
 This is above mentioned infrared-finite gauge
 $d_l=1-n=-1$. As  follows from eq.(\ref{Dcx}), in this gauge
 $$
 D_{\mu\nu}^c(x)=\frac{i}{2\pi}\frac{x_\mu x_\nu}{x^2-i0},
 $$
 and the equation for the electron propagator is simply
 \be
 S^{-1}=(S^c)^{-1}+2\alpha S.
 \label{S2}
 \ee
 In the momentum space eq.(\ref{S2}) is reduced to a system of
 algebraic equations, and among its solutions a DCSB-solution exists.
 Surely, in the two-dimensional theory the  spontaneous breaking
 of continuous chiral symmetry is not realized since  such a
 state is unstable in correspondence with Mermin-Wagner-Coleman
 theorem. Nevertheless, the fact itself is remarkable from
 the point of view of above mentioned analogy with four-dimensional QCD.\\

\section*{Acknowledgments}

Author is grateful to B~A~Arbuzov and P~A~Saponov for useful
discussions. The work is supported in part by RFBR, grant
No.98-02-16690.

\end{document}